\documentclass[nofootinbib,twocolumn,eqsecnum,superscriptaddress]{revtex4}
\usepackage{graphicx}
\usepackage{amsfonts}
\usepackage{amssymb}
\usepackage{amsmath}
\usepackage{bm}
\usepackage{dcolumn}
\usepackage{color}
\usepackage{ifthen,natbib,hyperref}

\newcommand{\be}{\begin{equation}}
\newcommand{\ee}{\end{equation}}
\newcommand{\ba}{\begin{eqnarray}}
\newcommand{\ea}{\end{eqnarray}}
\renewcommand{\d}{\mathrm{d}}

\begin{document}


\title{Gravitational waves in the black string braneworld}

\affiliation{Department of Mathematics and Statistics, University of
New Brunswick, Fredericton, NB, Canada~E3B~5A3}%
\affiliation{Cosmology \& Gravity Group, Department of Mathematics
and Applied Mathematics, University of Cape Town, Rondebosch 7701,
Cape Town, South Africa}%
\affiliation{Institute of Cosmology \& Gravitation, University of
Portsmouth, Portsmouth~PO1~2EG, UK}

\author{Sanjeev S.~Seahra}
\email{sseahra@unb.ca}%
\affiliation{Department of Mathematics and Statistics, University of
New Brunswick, Fredericton, NB, Canada~E3B~5A3}%
\affiliation{Institute of Cosmology \& Gravitation, University of
Portsmouth, Portsmouth~PO1~2EG, UK}

\author{Chris Clarkson}
\email{chris.clarkson@uct.ac.za}%
\affiliation{Cosmology \& Gravity Group, Department of Mathematics
and Applied Mathematics, University of Cape Town, Rondebosch 7701,
Cape Town, South Africa}%
\affiliation{Institute of Cosmology \& Gravitation, University of
Portsmouth, Portsmouth~PO1~2EG, UK}

\setlength\arraycolsep{2pt}

\newcommand*{\Y}{{Y}_{lm}}

\newcommand*{\Scalar}[1]{ {\mathsf{H}}_{#1} }
\newcommand*{\Tensor}{ {\mathsf{K}} }

\newcommand*{\dlangle}{\bm\{}
\newcommand*{\drangle}{\bm\}}

\newcommand*{\di}{\partial}
\newcommand*{\dimT}{\Theta}
\newcommand*{\bulk}{\text{bulk}}
\renewcommand{\d}{\mathrm{d}}

\date{\today}

\begin{abstract}

We study gravitational waves in the black string Randall-Sundrum
braneworld. We present a reasonably self-contained and complete
derivation of the equations governing the evolution of gravitational
perturbations in the presence of a brane localized source, and then
specialize to the case of spherical radiation from a pointlike body
in orbit around the black string.  We solve for the resulting
gravitational waveform numerically for a number of different orbital
parameters.

\end{abstract}

\maketitle

\section{Introduction}\label{sec:intro}

The Randall-Sundrum (RS) braneworld model postulates that our
universe is a 4-dimensional hypersurface embedded in 5-dimensional
space \cite{Randall:1999ee,Randall:1999vf,Maartens:2003tw}. One of
the most remarkable features of the model is that in the 10 years
since its introduction, no one has found any evidence of tension
with current observations or tests of gravitational phenomena.  This
is despite the fact that the model incorporates a large extra
dimension through which gravity is allowed to propagate. The RS
model's viability stems from the fact that it alters conventional
general relativity (GR) on scales smaller than the curvature scale
of the bulk spacetime $\ell$. Hence, if one chooses $\ell$ to be
sufficiently small the RS model is indistinguishable from GR in many
experimental or observational situations.

The principal virtue of the RS scenario is also a bit of a
detriment: In order to constrain or refute the model one has to look
at increasingly smaller scale phenomena.  The most direct test comes
from laboratory measurements of the gravitational force between two
masses, which yields $\ell \lesssim 50\,\mu\text{m}$
\cite{Adelberger:2003zx,Kapner:2006si}.  One can also examine high
energy cosmological phenomena to derive observable consequences of
the braneworld paradigm.  The idea is that when the Hubble radius
becomes smaller than the bulk curvature $H\ell \lesssim 1$, the
physical scale of all interesting gravitational interactions are
also smaller than $\ell$.  So in these epochs, one expects the RS
corrections to GR to become dominant.  The spectrum of tensor
perturbations in the high energy radiation RS era has been
calculated and shown to the be consistent with the GR result (with
minor modifications)
\cite{Hiramatsu:2004aa,Kobayashi:2005dd,Seahra:2006tm}. On the other
hand, the spectrum of scalar density perturbations is found to be
enhanced over the GR expectation in the early universe, which could
lead to an overproduction of primordial black holes in the RS model
\cite{Cardoso:2007zh}. The behaviour of scalar perturbations during
inflation has also been considered, and it was found that there were
very small corrections to the power spectrum of primordial
fluctuations \cite{Hiramatsu:2006cv,Koyama:2007as}.

\begin{figure}
\begin{center}
\includegraphics[width=0.7\columnwidth]{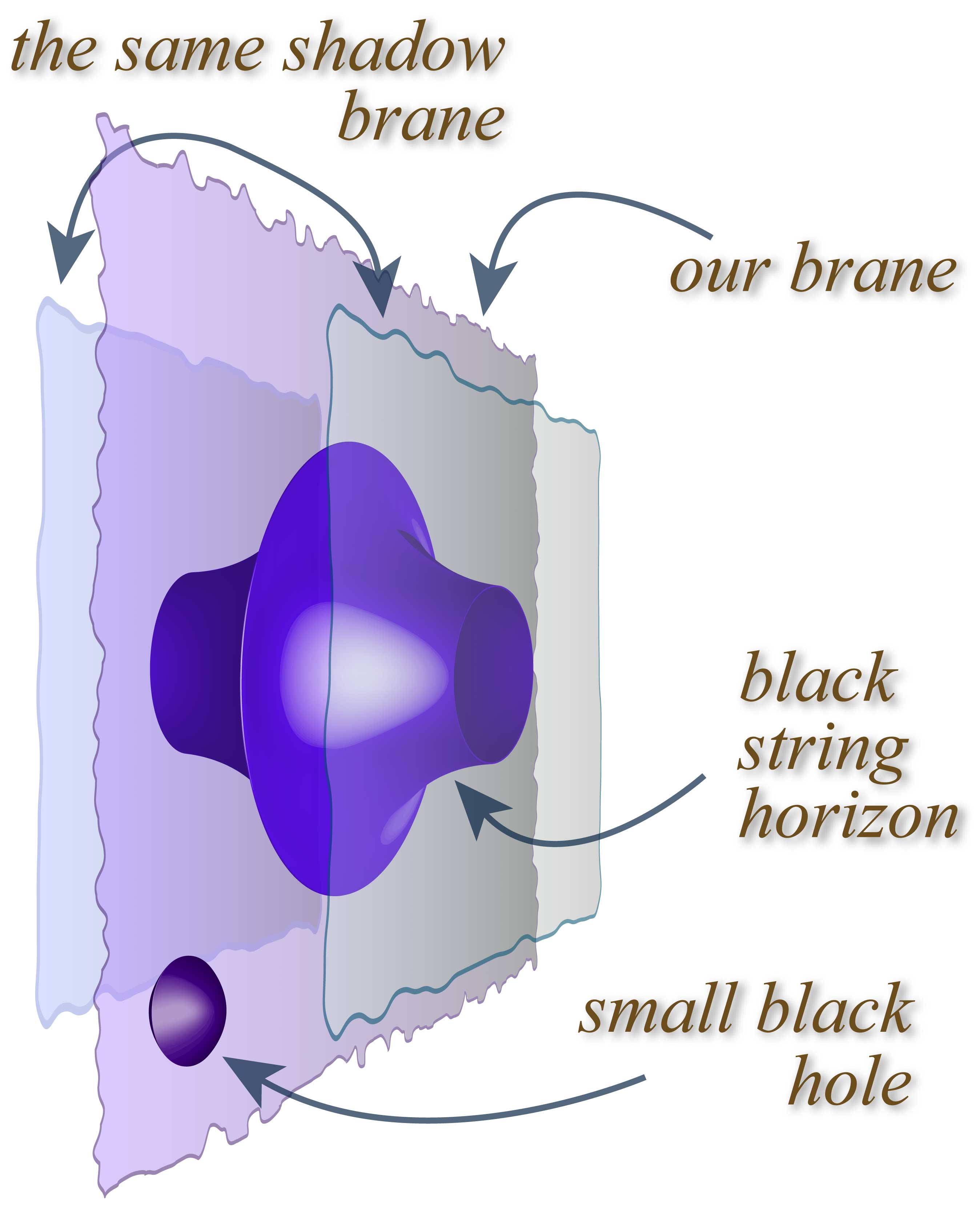}
\caption{A schematic of the black string braneworld. The black
string is a line singularity extending from our brane to a shadow
brane in the bulk.  There is a periodic identification such that the
background geometry is symmetric about our brane.  The singularity
is covered by a warped horizon, and the brane separation must be
large enough to be consistent with solar system tests of GR yet
small enough to avoid the Gregory-Laflamme instability
\cite{Gregory:1993vy,Gregory:2000gf,Seahra:2004fg}. In this paper,
we consider the situation where the string is perturbed by a small
orbiting body on one of the branes, here depicted as a small black
hole.} \label{fig:schematic}
\end{center}
\end{figure}
Another possible means of constraining the RS model is with
gravitational waves (GWs) with wavelengths $\lambda \lesssim \ell$.
These naturally probe gravitational interactions in the regime where
RS effects should be important, and can be viewed as the dynamical
counterpart to the static laboratory tests of Newton's Law mentioned
above. It is useful to have a concrete model of the generation and
propagation of these short-wavelength GWs in order to determine if
they have sufficient amplitude to be observed by real GW detectors.
To that end, we have previously considered the behaviour of
gravitational perturbations around a black string braneworld (see
Fig.~\ref{fig:schematic}) and showed that a generic feature of the
signal involves a long-lived oscillatory tail composes of a discrete
spectrum of modes whose wavelengths are less than $\ell$
\cite{Seahra:2004fg}. The amplitude of spherical radiation emitted
by a black string being perturbed by an orbiting small body was
estimated in \cite{Clarkson:2006pq}. This type of GW radiation is
considered to be a possible source for very-high frequency GW
detectors \cite{Cruise:2006zt,Nishizawa:2007tn,Nishizawa:2008se}.

Our purpose in this paper is to present the details of the black
string perturbative formalism utilized in the Letters
\cite{Seahra:2004fg,Clarkson:2006pq}.  This is the subject of
\S\ref{sec:RS}--\S\ref{sec:point}.  We also describe how to
numerically calculate the GWs sourced by a ``point particle''
orbiting the black string, and present the results of a number of
simulations in \S\ref{sec:typical}.

\section{A generalized Randall-Sundrum two brane
model}\label{sec:RS}

In this section, we present a generalized version of the
Randall-Sundrum two brane model in a coordinate invariant formalism.
Our treatment represents a generalization of the work of
\citet{Shiromizu:1999wj}. We begin by outlining the geometry of the
model, the action governing the dynamics, and the ensuing field
equations. We then specialize to the black string braneworld model,
which will be perturbed in the next section.

\subsection{Geometrical framework and notation}

Consider a (4+1)-dimensional manifold $(\mathcal{M},g)$, which we
refer to as the `bulk'. One of the spatial dimensions of
$\mathcal{M}$ is assumed to be compact; i.e., the 5-dimensional
topology is $\mathbb{R}^4 \times S$. We place coordinates $x^A$ on
$\mathcal{M}$ so that the 5-dimensional line element reads:
\begin{equation}
    ds_5^2 = g_{AB} dx^A dx^B.
\end{equation}
We assume that there is a scalar function $\Phi$ that uniquely maps
points in $\mathcal{M}$ into the interval $I = (-d,+d]$.  Here, $d$
is a constant parameter that is one of the fundamental length scales
of the problem.  The gradient of this mapping $\di_A\Phi$ satisfies
\begin{equation}
    \di_A \Phi \, \di^A \Phi > 0,
\end{equation}
and is tangent to the compact dimension of $\mathcal{M}$. This
scalar function defines a family of timelike hypersurfaces
$\Phi(x^A) = Y$, which we denote by $\Sigma_Y$.  The two
submanifolds at the endpoints of $I$, $\Sigma_{d}$ and
$\Sigma_{-d}$, are periodically identified.

Let us now place 4-dimensional coordinates $z^\alpha$ on each of
the $\Sigma_Y$ hypersurfaces.  These coordinates will be related
to their 5-dimensional counterparts by parametric equations of the
form: $x^A = x^A(z^\alpha)$.  We then define the following basis
vectors
\begin{gather}\nonumber
    e^A_\alpha = \frac{\di x^A}{\di z^\alpha}, \quad n^A =
    \frac{\di^A \Phi}{\sqrt{\di_B \Phi \, \di^B \Phi}},
    \\ \quad n_A e^A_\alpha = 0, \quad n^A n_A = +1. \label{basis
    vectors}
\end{gather}
The tetrad $e^A_\alpha$ is everywhere tangent to $\Sigma_Y$, while
$n^A$ is everywhere normal to $\Sigma_Y$.  The projection tensor
onto the $\Sigma_Y$ hypersurfaces is given by
\begin{equation}
    q_{AB} = g_{AB} - n_A n_B, \quad n^A q_{AB} = 0.
\end{equation}
From this, it follows that the intrinsic line element on each of
the $\Sigma_Y$ hypersurfaces is
\begin{equation}
    ds^2_4 = q_{\alpha\beta} dz^\alpha dz^\beta, \quad
    q_{\alpha\beta} = e^A_\alpha e^B_\beta q_{AB} = e^A_\alpha e^B_\beta g_{AB}.
\end{equation}
The object $q_{\alpha\beta}$ behaves as a tensor under
4-dimensional coordinate transformations $z^\alpha \rightarrow
\tilde{z}^\alpha(z^\beta)$ and is the induced metric on the
$\Sigma_Y$ hypersurfaces.  It has an inverse $q^{\alpha\beta}$
that can be used to define $e_A^\alpha$:
\begin{equation}
    e_A^\alpha = g_{AB} q^{\alpha\beta} e^B_\beta, \quad
    \delta^\alpha_\beta = q^{\alpha\gamma} q_{\gamma\beta} =
    e^\alpha_A e^A_\beta.
\end{equation}

Generally speaking, we define the projection of any 5-tensor
$T_{AB}$ onto the $\Sigma_Y$ hypersurfaces as
\begin{equation}
    T_{\alpha\beta} = e^A_\alpha e^B_\beta T_{AB},
\end{equation}
where the generalization to tensors of other ranks is obvious.  The
4-dimensional intrinsic covariant derivative of $T_{\alpha\beta}$ is
related to the 5-dimensional covariant derivative of $T_{AB}$ by
\begin{equation}
    [ \nabla_\alpha T_{\mu\nu} ]_q = e^A_\alpha e^M_\mu e^N_\nu \nabla_A
    q^B_M q^C_N T_{BC},
\end{equation}
where the notation $[\cdots]_q$ means that the quantity inside the
square brackets is calculated with the $q_{\alpha\beta}$ metric.

Finally, the extrinsic curvature of each $\Sigma_Y$ hypersurface
is:
\begin{gather}\nonumber
    K_{AB} = q^C_A \nabla_{C} n_B = \tfrac{1}{2} \pounds_n q_{AB} = K_{BA}, \quad n^A K_{AB} = 0,\\
    K_{\alpha\beta} = e^A_\alpha e^B_\beta K_{AB} = e^A_\alpha e^B_\beta \nabla_A
    n_B.
\end{gather}

\subsection{The action and field equations}

We label the hypersurfaces at $Y = y_+ = 0$ and $Y = y_- = +d$ as
the `visible brane' $\Sigma^+$ and `shadow brane' $\Sigma^-$,
respectively.  Our observable universe is supposed to reside on
the visible brane.  These hypersurfaces divide the bulk into two
halves: the lefthand portion $\mathcal{M}_\text{L}$ which has $y
\in (-d,0)$, and the righthand portion which has $y \in (0,+d)$.
The action for our model is:
\begin{eqnarray}
    \nonumber S & = & \frac{1}{2\kappa_5^2} \int\limits_{\mathcal{M}_\text{L}} \left[ {}^{(5)}R - 2
    \Lambda_5 \right] + \frac{1}{2\kappa_5^2} \int\limits_{\mathcal{M}_\text{R}} \left[ {}^{(5)}R - 2
    \Lambda_5 \right] \\ \nonumber & & +  \sum_{\epsilon=\pm} \frac{1}{2} \int\limits_{\Sigma^\epsilon}
    \left( \mathcal{L}^\epsilon - 2\lambda^\epsilon - \frac{1}{\kappa_5^2} [K]^\epsilon \right) \\
    & & + \frac{1}{2} \int\limits_{\mathcal{M}_\text{L}}
    \mathcal{L}_\text{L} + \frac{1}{2} \int\limits_{\mathcal{M}_\text{R}}
    \mathcal{L}_\text{R}.
\end{eqnarray}
In this expression, $\kappa_5^2$ is the 5-dimensional gravity
matter coupling, $\Lambda_5 = -6k^2$ is the bulk cosmological
constant, $\lambda^\pm = \pm 6k/\kappa_5^2$ are the brane
tensions, and $\ell = 1/k$ is the curvature length scale of the
bulk.  Also, $\mathcal{L}^\pm$ is the Lagrangian density of matter
residing on $\Sigma^\pm$, while $\mathcal{L}_\text{L}$ and
$\mathcal{L}_\text{R}$ are the Lagrangian densities of matter
living in the bulk.  Note that the visible brane in our model has
positive tension while the shadow brane has negative tension.

The quantity $[K]^\pm$ is the jump in the trace of the extrinsic
curvature of the $\Sigma_Y$ hypersurfaces across each brane.  To
clarify, suppose that $\di\mathcal{M}_\text{L}^\pm$ and
$\di\mathcal{M}_\text{R}^\pm$ are the boundaries of
$\mathcal{M}_L$ and $\mathcal{M}_R$ coinciding with $\Sigma^\pm$,
respectively. Then,
\begin{subequations}
\begin{eqnarray}
    [K]^+ & = & q^{\alpha\beta} K_{\alpha\beta}
    \Big|_{\di\mathcal{M}_\text{R}^+} - q^{\alpha\beta} K_{\alpha\beta}
    \Big|_{\di\mathcal{M}_\text{L}^+}, \\
    {[K]^-} & = & q^{\alpha\beta} K_{\alpha\beta}
    \Big|_{\di\mathcal{M}_\text{L}^-} - q^{\alpha\beta} K_{\alpha\beta}
    \Big|_{\di\mathcal{M}_\text{R}^-}.
\end{eqnarray}
\end{subequations}

We can now write down the field equations for our model.  Setting
the variation of $S$ with respect to the bulk metric $g^{AB}$
equal to zero yields that:
\begin{gather}\nonumber
    G_{AB} - 6k^2 g_{AB} = \kappa_5^2 \left[ \theta(+y)
    T^\text{R}_{AB} + \theta(-y) T^\text{L}_{AB} \right], \\
    \label{bulk field equation}
    T^\text{L,R}_{AB} = - \frac{2}{\sqrt{-g}} \frac{\delta \left( \sqrt{-g}
    \mathcal{L}_\text{L,R} \right)}{\delta g^{AB}}.
\end{gather}
Meanwhile, variation of $S$ with respect to the induced metric on
each boundary yields
\begin{multline}
    Q^\pm_{AB} = \left\{ [K_{AB}] \pm 2kq_{AB} + \kappa_5^2 (T_{AB} - \tfrac{1}{3} T q_{AB})
    \right\}^\pm = 0, \\ T_{AB}^\pm = e_A^\alpha e_B^\beta \left\{ -
    \frac{2}{\sqrt{-q}} \frac{\delta \left( \sqrt{-q}
    \mathcal{L} \right)}{\delta q^{\alpha\beta}}
    \right\}^\pm. \label{brane boundary conditions}
\end{multline}
Here, the $\{\cdots\}^\pm$ notation means that everything inside the
curly brackets is evaluated at $\Sigma^\pm$.  We see that (\ref{bulk
field equation}) are the bulk field equations to be satisfied by the
5-dimensional metric $g_{AB}$, while (\ref{brane boundary
conditions}) are the boundary conditions that must be enforced at
the position of each brane.  Of course, (\ref{brane boundary
conditions}) are simply the Israel junction conditions for thin
shells in general relativity.  In a braneworld context, the
symmetric versions of these equations first appeared in
\citet{Shiromizu:1999wj}.

In what sense is our model a generalization of the RS setup? The
original Randall-Sundrum model exhibited a $\mathbb{Z}_2$
symmetry, which implied that $\mathcal{M}_\text{L}$ is the mirror
image of $\mathcal{M}_\text{R}$.  Also, in the RS model the bulk
was explicitly empty.  However, since we allow for an asymmetric
distribution of matter in the bulk, we explicitly violate the
$\mathbb{Z}_2$ symmetry and bulk vacuum assumption.

\subsection{The black string braneworld}

We now introduce the black string braneworld, which is a
$\mathbb{Z}_2$ symmetric solution of (\ref{bulk field equation})
and (\ref{brane boundary conditions}) with no matter sources:
\begin{equation}
    \mathcal{L}_\text{L} \doteq \mathcal{L}_\text{R} \doteq \mathcal{L}^\pm
    \doteq 0.
\end{equation}
Here, we use $\doteq$ to indicate equalities that only hold in the
black string background.  The bulk geometry for this solution is
given by:
\begin{multline} \label{black string metric}
ds_5^2 \doteq a^2(y) \left[
-f(r)\,dt^2+\frac{1}{f(r)}\,dr^2+r^2\,d\Omega^2
\right] + dy^2, \\
f(r) = 1 - 2G M/r, \quad a(y) = {\rm e}^{-k|y|}.
\end{multline}
Here, $M$ is the mass parameter of the black string and $G =
\ell_\text{Pl}/M_\text{Pl}$ is the ordinary 4-dimensional Newton's
constant.  The function $\Phi$ used to locate the branes is
trivial in this background:
\begin{equation}
    \Phi(x^A) \doteq y,
\end{equation}
which means that the $\Sigma^\pm$ branes are located at $y = 0$
and $y = d$, respectively.  The $\Sigma_Y \doteq \Sigma_y$
hypersurfaces have the geometry of Schwarzschild black holes, and
there is 5-dimensional line-like curvature singularity at $r = 0$:
\begin{equation}
    R^{ABCD} R_{ABCD} \doteq \frac{48 G^2 M^2 e^{4k|y|}}{r^6} + 40 k^2.
\end{equation}
Note that the other singularities at $y = \pm\infty$ are excised
from our model by the restriction $y \doteq Y \in (-d,d]$, so we
will not consider them further.

Finally, note that the normal and extrinsic curvature associated
with the $\Sigma_Y$ hypersurfaces satisfy the following convenient
properties:
\begin{equation}
    n_A \doteq \di_A y, \quad n^A\nabla_A n^B \doteq 0, \quad K_{AB} \doteq -kq_{AB}.
\end{equation}
These expressions are used liberally below to simplify formulae
evaluated in the black string background.

\section{Linear perturbations}\label{sec:linear}

We now turn attention to perturbations of the black sting
braneworld.  Our treatment will be a reformulated and generalized
version of the original Randall-Sundrum work
\cite{Randall:1999ee,Randall:1999vf} and the seminal contribution of
\citet{Garriga:1999yh}.

\subsection{Perturbative variables}

We are ultimately interested in the behaviour of gravitational waves
in this model, which are described by fluctuations of the bulk
metric:
\begin{equation}
    g_{AB} \rightarrow g_{AB} + h_{AB},
\end{equation}
where $h_{AB}$ is understood to be a `small' quantity.  The
projection of $h_{AB}$ onto the visible brane is the observable that
can potentially be measured in gravitational wave detectors.  But it
is not sufficient to consider fluctuations in the bulk metric alone
--- to get a complete picture, we must also allow for the perturbation
of the matter content of the model as well as the positions of the
branes.

Obviously, matter perturbations are simply described by the
$T_{AB}^\text{L}$, $T_{AB}^\text{R}$, and $T_{AB}^\pm$ stress-energy
tensors, which are considered to be small quantities of the same
order as $h_{AB}$.  On the other hand, we describe fluctuations in
the brane positions via a perturbation of the scalar function
$\Phi$:
\begin{equation}
    \Phi(x^A) \rightarrow y + \xi(x^A).
\end{equation}
Here, $\xi$ is a small spacetime scalar.  Recall that the position
of each brane is implicitly defined by $\Phi(x^A) = y_\pm$.  Hence,
the brane locations after perturbation are given by the solution of
the following for $y$:
\begin{equation}
    y + \xi\Big|_{y = y_\pm} + (y - y_\pm) \di_y \xi\Big|_{y = y_\pm}  +
    \cdots = y_\pm.
\end{equation}
However, note that $y - y_\pm$ is of the same order as $\xi$, so at
the linear level the new brane positions are simply given by
\begin{equation}
    y = y_\pm - \xi\Big|_{y = y_\pm}.
\end{equation}
Hence, the perturbed brane positions are given by the brane bending
scalars:
\begin{equation}
    \xi^\pm = \xi\Big|_{y = y_\pm}, \quad n^A \di_A \xi^\pm = 0.
\end{equation}
Note that because $\xi^+$ and $\xi^-$ are explicitly evaluated at
the brane positions, they are essentially 4-dimensional scalars that
exhibit no dependence on the extra dimension.

Having now delineated a set of variables that parameterize the
fluctuations of the black string braneworld, we now need to
determine their equations of motion.

\subsection{Linearizing the bulk field equations}

First, we linearize the bulk field equations (\ref{bulk field
equation}) about the black string solution.  Notice that (\ref{bulk
field equation}) only depends on the bulk metric and the bulk matter
distribution.  Hence, the linearized field equations will only
involve $h_{AB}$, $T_{AB}^\text{L}$ and $T_{AB}^\text{R}$.  The
actual derivation of the equation proceeds in the same manner as in
4-dimensions, and we just quote the result:
\begin{multline}\label{general perturbation equation}
    \nabla^C \nabla_C h_{AB}  - \nabla^C
    \nabla_{A} h_{BC} - \nabla^C
    \nabla_{B} h_{AC} \\ +\nabla_A \nabla_B h^C{}_C - 8 k^2 h_{AB} =
     -2\kappa_5^2\Sigma_{AB}^\bulk,
\end{multline}
where
\begin{multline}
    \Sigma_{AB}^\bulk = \Theta(+y)(T^\text{R}_{AB}-\tfrac{1}{3}
    T^\text{R} g_{AB}) \\ +\Theta(-y)(T^\text{L}_{AB}-\tfrac{1}{3}
    T^\text{L} g_{AB}).
\end{multline}
The wave equation (\ref{general perturbation equation}) is valid
for arbitrary choices of gauge and generic matter sources. If we
specialize to the Randall-Sundrum gauge
\begin{equation}\label{TT conditions}
    \nabla^A h_{AB} = 0, \quad h^A{}_A = 0, \quad h_{AB} = e_A^\alpha e_B^\beta h_{\alpha\beta},
\end{equation}
eq.~(\ref{general perturbation equation}) reduces to
\begin{multline}\label{perturbation equation 3}
    \hat\Delta_{AB}{}^{CD} h_{CD} +
     (G Ma)^2(\pounds_n^2 - 4k^2) h_{AB} \\ =  -2(G Ma)^2\kappa_5^2 \Sigma_{AB}^\bulk,
\end{multline}
where we have defined the operator
\begin{align} \nonumber
    \hat\Delta_{AB}&{}^{CD} = (GMa)^2 [ q^{MN} \nabla_M q^P_N q^C_A q^D_B
    \nabla_P + 2 {}^{(4)}\!R_A{}^C{}_B{}^D ] \\ \nonumber = & \, (GMa)^2 e_A^\alpha e_B^\beta
    \Big[ \delta_\alpha^\gamma \delta_\beta^\delta \nabla^\rho
    \nabla_\rho + 2 {R}_\alpha\!{}^\gamma\!{}_\beta\!{}^\delta
    \Big]_q  e^C_\gamma e^D_\delta \\ = & \, (GM)^2 e_A^\alpha e_B^\beta
    \Big[ \delta_\alpha^\gamma \delta_\beta^\delta \nabla^\rho
    \nabla_\rho + 2 {R}_\alpha\!{}^\gamma\!{}_\beta\!{}^\delta
    \Big]_g e^C_\gamma e^D_\delta. \label{Delta operator def}
\end{align}
Here, ${}^{(4)}\!R_{ACBD}$ is the Riemann tensor on $\Sigma_y$,
which can be related to the 5-dimensional curvature tensor via the
Gauss equation
\begin{equation}
     {}^{(4)}\!R_{MNPQ} = q^A_M q^B_N q^C_P q^D_Q R_{ABCD}+ 2
     K_{M[P} K_{Q]N}.
\end{equation}
On the second line of (\ref{Delta operator def}) the 4-tensor
inside the square brackets is calculated using $q_{\alpha\beta}$.
We can re-express this object in terms of the ordinary
Schwarzschild metric $g_{\alpha\beta}$, which is conformally
related to $q_{\alpha\beta}$ via the warp factor:
\begin{subequations}
\begin{gather}
    q_{\alpha\beta} = a^2 g_{\alpha\beta}, \\
    g_{\alpha\beta} dz^\alpha dz^\beta = -f\,dt^2 + f^{-1}
    \, dr^2 + r^2 d\Omega^2.
\end{gather}
\end{subequations}
The quantity in square brackets on the third line of (\ref{Delta
operator def}) is calculated from
$g_{\alpha\beta}$.\footnote{Unless otherwise indicated, for the
rest of the paper any tensorial expression with Greek indices
should be evaluated using the Schwarzschild metric
$g_{\alpha\beta}$.}  One can easily confirm that
$\hat\Delta_{AB}{}^{CD}$ is `$y$-independent' in the sense that it
commutes with the Lie derivative in the $n^A$ direction:
\begin{equation}
    [{}^{(4)}\!\hat\Delta_{AB}{}^{CD},\pounds_n] = 0.
\end{equation}
In addition, the $(GM)^2$ prefactor makes $\hat\Delta_{AB}{}^{CD}$
dimensionless.

Notice that the lefthand side of (\ref{perturbation equation 3})
is both traceless and manifestly orthogonal to $n^A$, which
implies the following constraints on the bulk matter:
\begin{equation}
    \Sigma_{AB}^\bulk = e^\alpha_A e^\beta_B
    \Sigma_{\alpha\beta}^\bulk, \quad q^{\alpha\beta}
    \Sigma_{\alpha\beta}^\bulk = 0.
\end{equation}
In other words, our gauge choice is inconsistent with bulk matter
that violates these conditions.  If we wish to consider more
general bulk matter, we cannot use the Randall-Sundrum gauge.

\subsection{Linearizing the junction conditions}

Next, we consider the perturbation of the junction conditions
(\ref{brane boundary conditions}).  These can be re-written as
\begin{multline}\label{brane boundary condition 2}
    Q_{AB}^\pm =  \Big\{ [ \tfrac{1}{2} \nabla_{(A}n_{B)}-
    n_{(A|}n^C\nabla_C n_{|B)} ] \\ \pm k
    q_{AB} + \kappa_5^2 \left(
    T_{AB} - \tfrac{1}{3} T q_{AB} \right) \Big\}^\pm = 0.
\end{multline}
We require that $Q_{AB}^\pm$ vanish before and after perturbation,
so we need to enforce that the first order variation $\delta
Q_{AB}^\pm$ is equal to zero.

In order to calculate this variation, we can regard the tensors
$Q_{AB}^\pm$ as functionals the brane positions (as defined by
$\Phi$), the brane normals $n_A$, the bulk metric, and the brane
matter:
\begin{equation}
    Q_{AB}^\pm = Q_{AB}^\pm (\Phi,n_M,g_{MN},T^\pm_{MN}),
\end{equation}
from which it follows that
\begin{multline}\label{variation of Q}
    \delta Q_{AB}^\pm =  \left\{ \frac{\delta
    Q_{AB}}{\delta \Phi} \delta\Phi + \frac{\delta
    Q_{AB}}{\delta n_C} \delta n_C \right.  \\ \left.  + \frac{\delta
    Q_{AB}}{\delta g_{CD}} \delta g_{CD} + \frac{\delta
    Q_{AB}}{\delta T_{CD}} \delta T_{CD} \right\}_0^\pm.
\end{multline}
The $\{\cdots\}^\pm_0$ notation is meant to remind us that after we
have calculated the variational derivatives, we must evaluate the
expression in the background geometry at the \emph{unperturbed}
positions of the brane.

We now consider each term in (\ref{variation of Q}).  For
simplicity, we temporarily focus on the positive tension visible
brane and drop the + superscript. The first term represents the
variation of $Q_{AB}^\pm$ with brane position, which is covariantly
given by the Lie derivative in the normal direction:
\begin{equation}
    \left\{ \frac{\delta Q_{AB}}{\delta \Phi} \delta\Phi \right\}_0 = \left\{ -\xi \pounds_n
    Q_{AB} \right\}_0.
\end{equation}
But the Lie derivative of $Q_{AB}$ vanishes identically in the
background geometry, so this term is equal to zero.

The second term in (\ref{variation of Q}) represents the variation
of $Q_{AB}$ with respect to the normal vector.  Making note of the
definition (\ref{basis vectors}) of $n^A$ in terms of $\Phi$, as
well as $\delta\Phi = \xi$ and $n^A\nabla_A \xi = 0$, we arrive at
\begin{equation}
    \delta n_A = \nabla_A \xi, \quad n^A \delta n_A = 0.
\end{equation}
Notice that since the normal itself must be continuous across the
brane, we have $[ \delta n_A ] = 0$.  After some algebra, we find
that the variation of the junction conditions with respect to the
brane normal is non-zero and given by
\begin{equation}
    \left\{ \frac{\delta Q_{AB}}{\delta n_C} \delta n_C \right\}_0 =
    2q^C_A q^D_B \nabla_C \nabla_D \xi.
\end{equation}

The third term in (\ref{variation of Q}) is the variation with the
bulk metric itself $\delta g_{AB} = h_{AB}$. Calculating this is
straightforward, and the result is:
\begin{equation}
    \left\{ \frac{\delta Q_{AB}}{\delta g_{CD}} \delta g_{CD} \right\}_0 = \tfrac{1}{2}
    [ \pounds_n h_{AB} ] + 2k h_{AB}.
\end{equation}
The last variation we must consider is with respect to the brane
matter fields, which is trivial:
\begin{equation}
    \left\{ \frac{\delta Q_{AB}}{\delta T_{CD}} \delta T_{CD} \right\}_0 = \kappa_5^2 \left( T_{AB} -
    \tfrac{1}{3} T q_{AB} \right).
\end{equation}

So, we have the final result that
\begin{multline}\label{perturbed boundary conditions}
    \delta Q_{AB}^\pm = \left\{ 2q^C_A q^D_B \nabla_C \nabla_D \xi +
    \tfrac{1}{2} [ \pounds_n h_{AB} ]
    \right. \\ \left.
    \pm 2k h_{AB} + \kappa_5^2 \left( T_{AB} - \tfrac{1}{3} T q_{AB}
    \right) \right\}_0^\pm = 0.
\end{multline}
If we take the trace of $\delta Q_{AB}^\pm = 0$, we obtain
\begin{equation}\label{brane bending equations}
    q^{AB} \nabla_A \nabla_B \xi^{\pm} = \tfrac{1}{6} \kappa_5^2
    T^\pm.
\end{equation}
These are the equations of motion for the brane bending degrees of
freedom in our model, which are seen to be directly sourced by the
matter fields on each brane.

\subsection{Converting the boundary conditions into distributional sources}

We can incorporate the boundary conditions $\delta Q_{AB}^\pm = 0$
directly into the $h_{AB}$ equation of motion as delta-function
sources.  This is possible because the jump in the normal derivative
of $h_{AB}$ appears explicitly in the perturbed junction conditions.
This procedure gives
\begin{multline}\label{perturbation equation 4}
     \hat\Delta_{AB}{}^{CD} h_{CD} -
     \hat \mu^2 h_{AB} \\ =  -2(G Ma)^2\kappa_5^2 \left[ \Sigma_{AB}^\bulk +
     \sum_{\epsilon = \pm}
     \delta(y-y_\epsilon) \Sigma_{AB}^\epsilon \right].
\end{multline}
Here, we have defined
\begin{gather}\nonumber
    \hat\mu^2 = -(G Ma)^2 \left[ \pounds_n^2 +
    \frac{2\kappa_5^2}{3}  \sum_{\epsilon=\pm} \lambda^\epsilon \delta(y-y_\epsilon)
    - 4k^2 \right], \\
    \Sigma_{AB}^\pm = \left( T^\pm_{AB} - \tfrac{1}{3} T^\pm q_{AB}
    \right) + \frac{2}{\kappa_5^2} q^C_A q^D_B \nabla_C \nabla_D
    \xi^\pm.
\end{gather}
If we integrate the wave equation (\ref{perturbation equation 4})
over a small region traversing either brane, we recover the boundary
conditions (\ref{perturbed boundary conditions}).

Together with the gauge conditions,
\begin{equation}
    n^A h_{AB} = q^{AC} \nabla_A h_{CB} = 0 = q^{AB} h_{AB},
\end{equation}
(\ref{brane bending equations}) and (\ref{perturbation equation
4}) are the equations governing the perturbations of our model.

\section{Kaluza-Klein mode functions}\label{sec:KK}

\subsection{Separation of variables}

As mentioned above, we have that
\begin{equation}
    [\hat\Delta_{AB}{}^{CD},\pounds_n]h_{CD} = 0;
\end{equation}
i.e., $\hat\Delta_{AB}{}^{CD}$ is independent of $y$ when evaluated
in the $(t,r,\theta,\phi,y)$ coordinates.  This suggests that we
seek a solution for $h_{AB}$ of the form
\begin{equation}
    h_{AB} = Z \tilde{h}_{AB}, \quad \hat\mu^2 Z = \mu^2 Z,
\end{equation}
where,
\begin{equation}
    0 = \pounds_n \tilde{h}_{AB} \text{ and } 0 = q^{A}_B \nabla_A Z;
\end{equation}
that is, $Z$ is an eigenfunction of $\hat\mu^2$ with eigenvalue
$\mu^2$.  The existence of the delta functions in the $\hat\mu^2$
operator means that we need to treat the even and odd parity
solutions of this eigenvalue problem separately.

\subsection{Even parity eigenfunctions}

If $Z(-y) = Z(y)$, we see that $Z$ satisfies the following equations
in the interval $y \in [0,d]$:
\begin{gather}
\begin{split}
    m^2 Z(y) & = - a^2(y)(\di_y^2 - 4k^2)Z(y) , \\
    0 & = [(\di_y + 2k)Z(y)]_\pm, \\
    \mu & = GMm.
\end{split}
\end{gather}
There is a discrete spectrum of solutions to this eigenvalue problem
that are labeled by the positive integers $n = 1,2,3\ldots$:
\begin{multline}
    Z_n(y) = \alpha_n^{-1} [ Y_1(m_n\ell)J_2(m_n\ell e^{k|y|}) \\ - J_1(m_n\ell) Y_2(m_n\ell
    e^{k|y|}) ],
\end{multline}
where $\alpha_n$ is a constant, and $m_n = \mu_n/GM$ is the
$n^\text{th}$ solution of
\begin{equation}
    Y_1(m_n\ell)J_1(m_n\ell e^{kd}) = J_1(m_n\ell) Y_1(m_n\ell
    e^{kd}).
\end{equation}
There is also a solution corresponding to $m_0 = \mu_0 = 0$, which
is known as the zero-mode:
\begin{equation}
    Z_0(y) = \alpha_0^{-1} e^{-2k|y|}, \quad \alpha_0 = \sqrt{\ell}(1-e^{-2kd})^{1/2}.
\end{equation}
Hence, there exists a discrete set of solutions for bulk metric
perturbations of the form $h_{AB}^{(n)} = Z_n(y)
\tilde{h}_{AB}^{(n)}(z^\alpha)$.  When $n > 0$ these are called the
Kaluza-Klein (KK) modes of the modes, and the mass of any given mode
is given by the $m_n$ eigenvalue.  The $\alpha_n$ constants are
determined from demanding that $\{Z_n\}$ forms an orthonormal set
\begin{equation}
    \delta_{mn} = \int_{-d}^d dy \, a^{-2}(y) Z_m(y) Z_n(y).
\end{equation}
These basis functions then satisfy:
\begin{equation}\label{completeness}
    \delta(y - y_\pm) = \sum_{n=0}^\infty a^{-2} Z_n(y) Z_n(y_\pm).
\end{equation}
This identity is crucial to the model --- inspection of
(\ref{perturbation equation 4}) reveals that the brane stress energy
tensors appearing on the righthand side are multiplied by one of
$\delta(y - y_\pm)$.  Hence, brane matter only couples to the even
parity eigenmodes of $\hat\mu^2$.

\subsubsection*{Case 1: light modes}

It is useful to have simple approximate forms of the Kaluza-Klein
masses and normalization constants.  These are straightforward to
derive for modes that are `light' compared to mass scale set by the
AdS${}_5$ length parameter:
\begin{equation}
    m_n \ell \ll 1.
\end{equation}
Let us define a set of dimensionless numbers $x_n$ by:
\begin{equation}
    x_n = m_n \ell e^{kd}.
\end{equation}
Then for the light modes, we find that $x_n$ is the $n^\text{th}$
zero of the first-order Bessel function:
\begin{equation}\label{even masses}
    J_1(x_n) = 0.
\end{equation}
Also for light modes, the normalization constants reduce to
\begin{equation}
    \alpha_n \approx 2 \sqrt{\ell} \, e^{2kd}|J_0(x_n)|/\pi x_n, \quad n >
    0.
\end{equation}
Actually, it is more helpful to know the value of the KK mode
functions at the position of each brane.  We can parameterize these
as
\begin{equation}
    Z_n(y_\pm) = \sqrt{k} e^{-kd} z_n^\pm, \quad n>0.
\end{equation}
For the light Kaluza-Klein modes, the dimensionless $z_n^\pm$ are
given by
\begin{equation}\label{light z's}
    z_n^\pm \approx \left\{
    \genfrac{}{}{0pt}{0}{|J_0(x_n)|^{-1}}{e^{in\pi}} \right\}.
\end{equation}

\subsubsection*{Case 2: heavy modes}

At the other end of the spectrum, we have the heavy Kaluza-Klein
modes
\begin{equation}
    m_n \ell \gg 1.
\end{equation}
Under this assumption, we find
\begin{subequations}\label{heavy KK modes}
\begin{eqnarray}
    x_n & \approx & \frac{n\pi}{1-e^{-kd}},\\
    Z_n(y) & \approx & \sqrt{\frac{ke^{-k|y|}}{e^{kd}-1}} \cos \left[ n\pi
    \frac{e^{k|y|}-1}{e^{kd}-1} \right], \\
    z_n^\pm & \approx & \frac{1}{\sqrt{1-e^{-kd}}} \left\{
    \genfrac{}{}{0pt}{0}{e^{kd/2}}{e^{in\pi}} \right\}.
\end{eqnarray}
\end{subequations}
(Strictly speaking, an asymptotic analysis leads to formulae with
$n$ replaced by another integer $n'$ on the righthand sides of
Eqns.~(\ref{heavy KK modes}). However, we note that for even parity
modes, $n$ counts the number of zeroes of $Z_n(y)$ in the interval
$y \in (0,d)$, which allows us to deduce that $n' = n$.) Unlike the
analogous quantities for the light modes, $z_n^\pm$ shows an
explicit dependence on the dimensionless brane separation $d/\ell$.

\subsection{Odd parity eigenfunctions}

As mentioned above, brane matter only couples to Kaluza-Klein modes
with even parity.  But a complete perturbative description must
include the odd parity modes as well; for example, if we have matter
in the bulk distributed asymmetrically with respect to $y = 0$
(i.e.~$T_{AB}^\text{L} \ne T_{AB}^\text{R}$) modes of either parity
will be excited. Hence, for the sake of completeness, we list a few
properties of the odd parity Kaluza-Klein modes here.

Assuming $Z(-y) = -Z(y)$, we have:
\begin{gather}
\begin{split}
    m^2 Z(y) & = - a^2(y)(\di_y^2 - 4k^2)Z(y) , \\
    0 & = Z(y_+) = Z(y_-).
\end{split}
\end{gather}
Again, we have a discrete spectrum of solutions, this time labeled
by half integers:
\begin{multline}
    Z_{n+\frac{1}{2}}(y) = \alpha_{n+\frac{1}{2}}^{-1} [ Y_2(m_{n+\frac{1}{2}}\ell)
    J_2(m_{n+\frac{1}{2}}\ell e^{k|y|}) \\ - J_2(m_{n+\frac{1}{2}}\ell) Y_2(m_{n+\frac{1}{2}}\ell
    e^{k|y|}) ].
\end{multline}
The mass eigenvalues are now the solutions of
\begin{multline}
    Y_2(m_{n+\frac{1}{2}}\ell) J_2(m_{n+\frac{1}{2}}\ell e^{kd}) =
    \\ J_2(m_{n+\frac{1}{2}}\ell) Y_2(m_{n+\frac{1}{2}}\ell e^{kd}).
\end{multline}
Proceeding as before, we define
\begin{equation}
    x_{n+\frac{1}{2}} = m_{n+\frac{1}{2}}\ell e^{kd}.
\end{equation}
For light modes with $m_{n+\frac{1}{2}}\ell \ll 1$,
$x_{n+\frac{1}{2}}$ is the $n^\text{th}$ zero of the second-order
Bessel function:
\begin{equation}\label{odd masses}
    J_2(x_{n+\frac{1}{2}}) = 0.
\end{equation}
Taken together, (\ref{even masses}) and (\ref{odd masses}) imply the
following for the light modes:
\begin{equation}
    m_1 < m_{3/2} < m_2 < m_{5/2} < \cdots ;
\end{equation}
i.e., the first odd mode is heavier than the first even mode, etc.

Finally, we note that since the odd modes vanish at the background
position of the visible brane, it is impossible for us to observe
them directly within the context of linear theory.  This can change
at second order, since brane bending can allow us to directly sample
regions of the bulk where $Z_{n+\frac{1}{2}} \ne 0$.  However, this
phenomenon is clearly beyond the scope of this paper.

\subsection{Stability criterion}

Finally, as discussed in detail elsewhere \cite{Seahra:2004fg}, the
black string braneworld will be perturbatively stable if the
smallest KK mass satisfies
\begin{equation}
    \mu_1 = G Mm_1 > \mu_c \approx 0.4301.
\end{equation}
Under the approximation that the first mode is light ($x_1 e^{-kd}
\ll 1$) and using $G = \ell_\text{Pl}/M_\text{Pl}$, this gives a
restriction on the black string mass
\begin{equation}
    \frac{M}{M_\text{Pl}} \gtrsim \frac{\ell}{\ell_\text{Pl}}
    \frac{\mu_c}{x_1} e^{kd},
\end{equation}
or equivalently,
\begin{equation}
    \frac{M}{M_\odot}
    \gtrsim 8 \times 10^{-9} \left( \frac{\ell}{\text{0.1 mm}} \right) e^{d/\ell}.
\end{equation}
If we take $\ell =$ 0.1 mm, then we see that all solar mass black
holes will in actuality be stable black strings provided that
$d/\ell \lesssim 19$.

\section{Recovering 4-dimensional gravity}\label{sec:brans-dicke}

Let us now describe the limit in which we recover general
relativity.  (\citet{Garriga:1999yh} first considered this problem
in Minkowski space, but the approach employed here is somewhat
different.) We assume there are no matter perturbations in the bulk
and on the hidden brane; hence, we may consistently neglect the odd
parity Kaluza-Klein modes. By virtue of the brane bending equation
of motion (\ref{brane bending equations}), we can consistently set
$\xi^- = 0$. Furthermore, (\ref{completeness}) can be used to
replace the delta function in front of $\Sigma_{AB}^+$ in equation
(\ref{perturbation equation 4}).  We obtain,
\begin{multline}\label{perturbation equation 5}
     \hat\Delta_{AB}{}^{CD} h_{CD} -
     \hat \mu^2 h_{AB} \\ =  -2(G M)^2\kappa_5^2
     \Sigma_{AB}^+ \sum_{n=0}^\infty Z_n(y_+) Z_n(y).
\end{multline}
We now note that for $e^{-kd} \ll 1$,
\begin{equation}
    Z_0(y_+) = \sqrt{k}(1-e^{-2kd})^{-1/2} \gg Z_n(y_+), \quad n > 0.
\end{equation}
That is, the $n > 0$ terms in the sum are much smaller than the
$0^\text{th}$ order contribution.  This motivates an approximation
where the $n>0$ terms on the righthand side of (\ref{perturbation
equation 5}) are neglected, which is the so-called `zero-mode
truncation'.

When this approximation is enforced, we find that $h_{AB}$ must be
proportional to $Z_0(y)$; i.e., there is no contribution to $h_{AB}$
from any of the KK modes.  Hence, we have $\hat\mu^2 h_{AB} = 0$.
The resulting expression has trivial $y$ dependence, so we can
freely set $y = y_+$ to obtain the equation of motion for $h_{AB}$
at the \emph{unperturbed} position of the visible brane:
\begin{equation}\label{perturbation equation 6}
     \hat\Delta_{AB}{}^{CD} h^+_{CD}
     = -2(G M)^2\kappa_5^2
     \Sigma_{AB}^+ Z^2_0(y_+)
\end{equation}
But we are not really interested in $h_{AB}^+$, the physically
relevant quantity is the perturbation of the induced metric on the
perturbed brane, which is defined as the variation of
\begin{equation}
    q_{AB}^+ = [ g_{AB} - n_A n_B ]^+.
\end{equation}
We calculate $\delta q^+_{AB}$ in the same way as we calculated
$\delta Q_{AB}^\pm$ above (except for the fact that $q_{AB}$ shows
no explicit dependence on $T_{AB}^+$):
\begin{equation}
    \delta q_{AB}^+ = \left\{ \frac{\delta
    q_{AB}}{\delta \Phi} \delta\Phi + \frac{\delta
    q_{AB}}{\delta n_C} \delta n_C + \frac{\delta
    q_{AB}}{\delta g_{CD}} \delta g_{CD} \right\}_0^+.
\end{equation}
These variations are straightforward, and we obtain:
\begin{multline}\label{brane metric}
    \delta q_{AB}^+ \equiv \bar{h}_{AB}^+ = h^+_{AB} + 2k\xi^+
    q_{AB}^+ \\ - (n_A \nabla_B + n_B \nabla_A) \xi^+,
\end{multline}
where all quantities on the right are evaluated in the background
and at the unperturbed position of the brane.  Note that
$\bar{h}_{AB} n^A \ne 0$, which reflects the fact that $n_A$ is no
longer the normal to the brane after perturbation.

We now define the 4-tensors
\begin{equation}
    \bar{h}_{\alpha\beta}^+ = e^A_\alpha e^B_\beta \bar{h}_{AB}^+, \quad
    T^+_{\alpha\beta} = e^A_\alpha e^B_\beta T^+_{AB}.
\end{equation}
Here, $\bar{h}_{\alpha\beta}^+$ is the actual metric perturbation on
the visible brane.  Note that this perturbation is neither
transverse or tracefree:
\begin{equation}
    \nabla^\gamma \bar{h}^+_{\gamma \alpha} = 2k \nabla_\alpha \xi^+,
    \quad g^{\alpha\beta} \bar{h}^+_{\alpha\beta} = 8k\xi^+.
\end{equation}
We can now re-express the equation of motion (\ref{perturbation
equation 6}) in terms of $\bar{h}^+_{\alpha\beta}$ instead of
$h_{AB}^+$ using (\ref{brane metric}).  Dropping the $+$
superscripts, we obtain
\begin{multline}
    \nabla^\gamma \nabla_\gamma \bar{h}_{\alpha\beta}
    + \nabla_\alpha \nabla_\beta \bar{h}^\gamma\!{}_\gamma -
    \nabla^\gamma \nabla_\alpha \bar{h}_{\beta\gamma} -
    \nabla^\gamma \nabla_\beta \bar{h}_{\alpha\gamma}
    = \\ -2 Z_+^2 \kappa_5^2 \left[ T_{\alpha\beta} -
    \frac{1}{3} \left( 1 + \frac{k}{2Z^2_+} \right)
    T^\gamma\!{}_{\gamma} g_{\alpha\beta} \right] \\ + ( 6k - 4Z_+^2 )
    \nabla_\alpha \nabla_\beta \xi,
\end{multline}
where we have defined
\begin{equation}
    Z_+^2 = Z^2_0(y_+) = k (1 - e^{-2kd})^{-1}.
\end{equation}
In obtaining this expression, we have made use of the $\xi$ equation
of motion:
\begin{equation}
    g^{\alpha\beta} \nabla_{\alpha} \nabla_\beta \xi = \tfrac{1}{6}
    \kappa_5^2 g^{\alpha\beta} T_{\alpha\beta}.
\end{equation}

Note that we still have the freedom to make a gauge transformation
on the brane that involves an arbitrary 4-dimensional coordinate
transformation generated by $\eta_\alpha$:
\begin{equation}
    \bar{h}_{\alpha\beta} \rightarrow \bar{h}_{\alpha\beta} +
    \nabla_\alpha \eta_\beta + \nabla_\beta \eta_\alpha.
\end{equation}
We can use this gauge freedom to impose the condition
\begin{equation}
    \nabla_\beta \bar{h}^\beta\!{}_\alpha - \tfrac{1}{2}
    \nabla_\alpha \bar{h}^\beta\!{}_\beta = (2Z_+^2 - 3k)
    \nabla_\alpha \xi.
\end{equation}
Then, the equation of motion for 4-metric fluctuations reads
\begin{multline}\label{Brans-Dicke}
    \nabla^\gamma \nabla_\gamma \bar{h}_{\alpha\beta} +
    2R_\alpha\!{}^\gamma\!{}_\beta\!{}^\delta \bar{h}_{\gamma\delta} = \\
    -16\pi G \left[ T_{\alpha\beta} -
    \left( \frac{1+\omega_\text{\tiny BD}}{3+2\omega_\text{\tiny BD}} \right)
    T^\gamma\!{}_{\gamma} g_{\alpha\beta} \right],
\end{multline}
where we have identified
\begin{equation}
    \omega_\text{\tiny BD} = \frac{3}{2}(e^{2d/\ell} - 1), \quad
    G = \frac{\kappa_5^2}{8\pi\ell(1-e^{-2d/\ell})}.
\end{equation}
We see that (\ref{Brans-Dicke}) matches the equation governing
gravitational waves in a Brans-Dicke theory with parameter
$\omega_\text{\tiny BD}$.  Hence in the zero-mode truncation, the
perturbations of the black string braneworld are indistinguishable
from a 4-dimensional scalar tensor theory.

Note that (\ref{Brans-Dicke}) must hold everywhere in our model, so
we can consider the situation where our solar system is the
perturbative brane matter located somewhere in the extreme far-field
region of the black string.  The forces between the various
celestial bodies will be governed by (\ref{Brans-Dicke}) in the
$R_{\alpha\beta\gamma\delta} \approx 0$ limit.  In this scenario,
solar system tests of general relativity \cite{Will:2005va} place
bounds on the Brans-Dicke parameter, and hence $d/\ell$:
\begin{equation}
    \omega_\text{\tiny BD} \gtrsim 4 \times 10^4 \quad \Rightarrow \quad d/\ell
    \gtrsim 5.
\end{equation}
This lower bound on the dimensionless brane separation will be an
important factor in the discussion below.

\section{Spherical waves on the brane}\label{sec:spherical}

In this section, we specialize to the situation where there is
perturbative matter located on one of the branes and no other
sources.  Unlike Sec.~\ref{sec:brans-dicke}, our interest here is to
predict deviations from general relativity, so we will not use the
zero-mode truncation.  Principally for reasons of simplicity, we
will focus on spherically symmetric radiation, which is a channel
unavailable in the standard 4-dimensional setup.

\subsection{Mode decomposition}

To begin, we make the assumptions
\begin{equation}
    \Sigma^\text{bulk}_{AB} = 0, \text{ and } \Sigma_{AB}^+ = 0
    \text{ or } \Sigma_{AB}^- = 0;
\end{equation}
i.e., we set the matter perturbation in the bulk and one of the
branes equal to zero.  Note that due to the linearity of the problem
we can always add up solutions corresponding to different types of
sources; hence, if we had a physical situation with many different
types of matter, it would be acceptable to solve for the radiation
pattern induced by each source separately and then sum the results.

We decompose $h_{AB}$ as
\begin{equation}\label{main mode decomposition}
    h_{AB} = \frac{\kappa_5^2 (G M)^2 }{\mathcal{C}} e_A^\alpha e_B^\beta \sum_{n=0}^\infty Z_n(y)
    Z_n(y_\pm) h_{\alpha\beta}^{(n)}.
\end{equation}
Here, $\mathcal{C}$ is a normalization constant (to be specified
later) with dimensions of $(\text{mass})^{-4}$, and the expansion
coefficients $h_{\alpha\beta}^{(n)}$ are dimensionless.  We define a
dimensionless brane stress-energy tensors and brane bending scalars
by
\begin{equation}\label{Theta defn}
    \dimT^\pm_{\alpha\beta} = \mathcal{C} e^A_\alpha e^B_\beta T^\pm_{AB},
    \quad \tilde\xi^\pm = \frac{\mathcal{C}\xi^\pm}{(G M)^2\kappa_5^2}.
\end{equation}
Omitting the $\pm$ superscripts, we find that the equation of motion
for $h_{\alpha\beta}^{(n)}$ is
\begin{multline}\label{4D EOM 1}
    (GM)^2 \left[ \nabla^\gamma \nabla_\gamma h^{(n)}_{\alpha\beta} + 2R_\alpha\!{}^\gamma\!{}_\beta\!{}^\delta
    h^{(n)}_{\gamma\delta} \right] - \mu_n^2 h^{(n)}_{\alpha\beta} = \\
    -2 \left( \dimT_{\alpha\beta} - \tfrac{1}{3} \dimT
    g_{\alpha\beta} \right) - 4(GM)^2 \nabla_\alpha \nabla_\beta
    \tilde\xi,
\end{multline}
while the equation of motion for $\tilde\xi$ is
\begin{equation}\label{4D EOM 2}
    \nabla^\alpha \nabla_\alpha \tilde\xi = \tfrac{1}{6} \dimT.
\end{equation}
We also have the conditions
\begin{equation}
    \nabla^\alpha h_{\alpha\beta}^{(n)} = \nabla^\alpha
    \dimT_{\alpha\beta} = 0 = g^{\alpha\beta}
    h_{\alpha\beta}^{(n)}.
\end{equation}
Note that in all of these equations, all 4-dimensional quantities
are to be calculated with the Schwarzschild metric
$g_{\alpha\beta}$.  In particular, $\dimT = g^{\alpha\beta}
\dimT_{\alpha\beta}$.

\subsection{The radiative $s$-wave channel}

We now turn our attention to solving the coupled system of equations
(\ref{4D EOM 1}) and (\ref{4D EOM 2}) for a generic source
$\dimT_{\alpha\beta}$.  The symmetry of the background geometry
dictates that we decompose the problem in terms of spherical
harmonics:
\begin{subequations}\label{decompositions}
\begin{eqnarray}
  \tilde{\xi} & = & \frac{\xi^{(s)}}{\sqrt{4\pi}} +
  \sum_{l=1}^\infty \sum_{m=-l}^l Y_{lm} \tilde\xi_{lm}, \\
  h^{(n)}_{\alpha\beta} & = & \frac{h_{\alpha\beta}^{(n,s)}}{\sqrt{4\pi}}
  + \sum_{l=1}^\infty \sum_{m=-l}^l \sum_{i=1}^{10} \,
  [ Y^{(i)}_{lm} ]_{\alpha\beta} \, h^{(nlm)}_{i} \!\! , \\
  \label{Theta decomposition} \dimT_{\alpha\beta} & = &
  \frac{\dimT^{(s)}_{\alpha\beta}}{\sqrt{4\pi}} +
  \sum_{l=1}^\infty \sum_{m=-l}^l \sum_{i=1}^{10} \,
  [ Y^{(i)}_{lm} ]_{\alpha\beta} \, \dimT^{(lm)}_{i}.
\end{eqnarray}
\end{subequations}
Here, $[ Y^{(i)}_{lm}]_{\alpha\beta}$ are the tensorial spherical
harmonics in 4 dimensions.  The terms with $l > 0$ in this
decomposition can be quite involved, so for the purposes of this
paper we concentration on the spherically symmetric $s$-wave ($l=0$)
sector, which is described by $\xi^{(s)}$,
$h_{\alpha\beta}^{(n,s)}$, and $\dimT^{(s)}_{\alpha\beta}$.

We write the $l = 0$ contribution to the metric perturbation as
\begin{equation}\label{spherical h decomp}
    h^{(n,s)}_{\alpha\beta} = \Scalar{1} \, t_\alpha t_\beta - 2 \Scalar{2} \, t_{(\alpha}
    r_{\beta)} +
    \Scalar{3} \, r_\alpha r_\alpha + \Tensor \, \gamma_{\alpha\beta},
\end{equation}
where we have defined the orthonormal vectors
\begin{equation}
    t^\alpha = f^{-1/2} \di_t, \quad r^\alpha = f^{1/2} \di_r,
\end{equation}
which are pointing in the time and radial directions, respectively;
and
\begin{equation}
    \gamma_{\alpha\beta} = g_{\alpha\beta}+t_\alpha t_\beta -
    r_\alpha r_\beta, \quad t^\alpha \gamma_{\alpha\beta} = r^\alpha
    \gamma_{\alpha\beta} = 0.
\end{equation}
which is the induced metric on the 2-spheres of constant $r$ and
$t$.   Each of the expansion coefficients is a function of $t$ and
$r$; i.e., $\Scalar{i} = \Scalar{i}(t,r)$ and $\Tensor =
\Tensor(t,r)$. Notice that the condition that ${}^0
h^{(n)}_{\alpha\beta}$ is tracefree implies
\begin{equation}
    \Tensor = \tfrac{1}{2}( \Scalar{1} - \Scalar{3} ).
\end{equation}

Before going further, it is useful to define dimensionless
coordinates:
\begin{equation}
    \rho = \frac{r}{GM}, \quad \tau = \frac{t}{GM}, \quad x =
    \rho + 2 \ln\left(\frac{\rho}{2}-1 \right).
\end{equation}
Then, when our decompositions (\ref{decompositions}) are substituted
into the equations of motion, we find that all components of the
metric perturbation are governed by master variables
\begin{equation}\label{master variable def}
    \psi = \frac{2\rho^3}{2+\mu^2\rho^3} \left( \rho \frac{\di \Tensor}{\di\tau} - f
    \Scalar{2} \right), \quad \varphi = \rho \frac{\di \xi^{(s)}}{\di\tau}.
\end{equation}
Both $\psi = \psi(\tau,x)$ and $\varphi = \varphi(\tau,x)$ satisfy
simple wave equations:
\begin{subequations}\label{wave equations}\label{eq:master}
\begin{eqnarray}
  \label{psi equation} (\di_\tau^2 - \di_x^2 + V_\psi) \psi & = & \mathcal{S}_\psi + \mathcal{\hat{I}}\,\varphi, \\
  \label{phi equation} (\di_\tau^2 - \di_x^2 + V_\varphi) \varphi & = & \mathcal{S}_\varphi.
\end{eqnarray}
\end{subequations}
The potential and matter source term in the $\psi$ equation are:
\begin{subequations}
\begin{gather}
\begin{split}
    V_\psi = \frac{f}{{{\rho}^{3} \left(
                 2+{\rho}^{3}{\mu}^{2} \right)^{2}}} & \Big[
                 {\mu}^{6} {\rho}^{9} +6\,{\mu}^{4}{
                 \rho}^{7}-18\,{\mu}^{4}{\rho}^{6} \\ &
                 -24\,{\mu}^{2}
                 {\rho}^{4}+36\,{\mu}^{2}{\rho }^{3}+8 \Big],
\end{split} \\
\begin{split}
    \mathcal{S}_\psi = \frac{2f\rho^3  }{3(2+\mu^2\rho^3)^2}
    \Big[ \rho & (2+\mu^2  \rho^3 )
    \di_\tau \! \left( 2\Lambda_1 +3 \Lambda_3 \right)
     \\  & + 6(\mu^2\rho^3-4) f \Lambda_2
    \Big].
\end{split}
\end{gather}
\end{subequations}
Here, we have defined the following three scalars derived from the
dimensionless stress-energy tensor $\dimT^{(s)}_{\alpha\beta}$:
\begin{subequations}\label{Lambda defs}
\begin{eqnarray}
    \Lambda_1 & = & -\dimT^{(s)}_{\alpha\beta} g^{\alpha\beta}, \\
    \Lambda_2 & = & -\dimT_{\alpha\beta}^{(s)} t^\alpha r^\beta, \\
    \Lambda_3 & = & +\dimT_{\alpha\beta}^{(s)} \gamma^{\alpha\beta}.
\end{eqnarray}
\end{subequations}
The potential and source terms in the brane-bending equation are
somewhat less involved:
\begin{equation}
    V_\varphi = \frac{2f}{\rho^3}, \quad \mathcal{S}_\varphi = \frac{\rho f}{6} \di_\tau \Lambda_1.
\end{equation}
Finally, the interaction operator is
\begin{equation}
    \mathcal{\hat{I}} = \frac{8f}{(2+\mu^2\rho^3)^2} \left[6f\rho^2
    \di_\rho  + (\mu^2\rho^3 - 6\rho + 8) \right].
\end{equation}

\subsection{Inversion formulae}

Assuming that we can solve the wave equations (\ref{wave equations})
for a given source, we need formulae that allow us to express
$\Scalar{i}$, $\Tensor$ in terms of $\psi$ and $\varphi$ in order to
make gravitational wave prediction. This can be derived by inverting
the master variable definitions (\ref{master variable def}) with the
aid (\ref{wave equations}). The general formulae are actually very
complicated and not particularly enlightening, so we do not
reproduce them here. Ultimately, to make observational predictions
it is sufficient to know the form of the metric perturbation far
away from the black string and the matter sources, so we evaluate
the general inversion formulae in the limit of $\rho \rightarrow
\infty$ and with $\Lambda_i = 0$:
\begin{eqnarray}\nonumber
    \di_\tau \Scalar{1} & = & \frac{1}{\rho} \left[ \left( \di_\tau^2 +
    \frac{3}{\rho} \di_\rho + \mu^2 \right) \psi + \frac{4}{\mu^2} \left(
    \di_\tau^2 - \frac{1}{\rho} \di_\rho \right) \varphi
    \right], \\ \nonumber \Scalar{2} & = & \frac{1}{\rho} \left[ \left(
    \di_\rho + \frac{2}{\rho} \right) \psi + \frac{4}{\mu^2} \left(
    \di_\rho - \frac{1}{\rho} \right) \varphi \right], \\ \nonumber \di_\tau\Scalar{3}
    & = & \frac{1}{\rho} \left[ \left( \di_\tau^2 + \frac{1}{\rho}
    \di_\rho \right) \psi + \frac{4}{\mu^2} \left( \di_\tau^2 -
    \frac{2}{\rho} \di_\rho \right) \varphi \right], \\
    \di_\tau \Tensor & = & \frac{1}{\rho} \left[ \left(
    \frac{1}{\rho} \di_\rho + \frac{\mu^2}{2} \right) \psi +
    \frac{4}{\mu^2\rho} \left( \di_\rho - \frac{1}{\rho} \right)
    \varphi \right].\label{inversion formulae}
\end{eqnarray}
Note that these do not actually complete the inversion; in most
cases, a quadrature is also required to arrive at the final form of
the metric perturbation.

\section{Point particle sources on the brane}\label{sec:point}

We now specialize to the situation where the perturbing brane matter
is a ``point particle'' located on one of the branes.  We take the
particle Lagrangian density to be
\begin{equation}
    \mathcal{L}_p^\pm = \frac{M_p}{2} \left\{ \int \frac{\delta^4(z^\mu - z_p^{\mu})}
    {\sqrt{-q}} q_{\alpha\beta} \frac{dz_p^\alpha}{d\eta} \frac{dz_p^\beta}{d\eta}
    d\eta \right\}^\pm.
\end{equation}
In this expression, $\eta$ is a parameter along the particle's
trajectory as defined by the $q_{\alpha\beta}$ metric, $z_p^\mu$ are
the 4 functions describing the particle's position on the brane, and
$M_p$ is the particle's mass parameter.  Using (\ref{brane boundary
conditions}, we find the stress-energy tensor
\begin{equation}
    T^\pm_{\alpha\beta} = M_p \left\{ \int \frac{\delta^4(z^\mu - z_p^{\mu})}
    {\sqrt{-q}} q_{\alpha\rho} q_{\beta\lambda} \frac{dz_p^\rho}{d\eta} \frac{dz_p^\lambda}{d\eta}
    d\eta \right\}^\pm.
\end{equation}
The contribution from the particle to the total action is
\begin{equation}
    S_p^\pm = \frac{1}{2} \int\limits_{\Sigma^\pm} \mathcal{L}_p^\pm =
    \frac{M_p}{4} \int q^\pm_{\alpha\beta} \frac{dz_p^\alpha}{d\eta} \frac{dz_p^\beta}{d\eta}
    d\eta.
\end{equation}
Varying this with respect to the trajectory $z_p^\alpha$ and
demanding that $\eta$ is an affine parameter yields that the
particle follows a geodesic along the brane:
\begin{equation}
    \frac{d^2 z_p^\alpha}{d\eta^2} +
    \Gamma^\alpha_{\beta\gamma}[q^\pm]
    \frac{dz_p^\beta}{d\eta} \frac{dz_p^\gamma}{d\eta} = 0, \quad
    -1 = q^\pm_{\alpha\beta} \frac{dz_p^\alpha}{d\eta}
    \frac{dz_p^\beta}{d\eta},
\end{equation}
where $\Gamma^\alpha_{\beta\gamma}[q^\pm]$ are the Christoffel
symbols defined with respect to the $q_{\alpha\beta}^\pm$ metric.

We note that the above formulae make explicit use of the induced
brane metrics $q_{\alpha\beta}^\pm$.  However, all of our
perturbative formalism is in terms of the Schwarzschild metric
$g_{\alpha\beta}$, especially the definition of the $\Lambda_i$
scalars (\ref{Lambda defs}).  Hence, it is useful to translate the
above expressions using the following definitions:
\begin{equation}
    \eta = a_\pm \lambda, \quad u^\alpha =
    \frac{dz_p^\alpha}{d\lambda}, \quad -1 = g_{\alpha\beta}
    u^\alpha u^\beta.
\end{equation}
Then, the stress-energy tensor and particle equation of motion
become
\begin{equation}
    T^\pm_{\alpha\beta} = \frac{M_p}{a_\pm} \int \frac{\delta^4(z^\mu - z_p^{\mu})}
    {\sqrt{-g}} u_\alpha u_\beta \,d\lambda,\quad u^\alpha
    \nabla_\alpha u^\beta = 0.
\end{equation}
Note that the only difference between the stress-energy tensors on
the positive and negative tension branes is an overall division by
the warp factor.

By switching over to dimensionless coordinates, transforming the
integration variable to $\tau$ from $\lambda$, and making use of the
spherical harmonic completeness relationship, we obtain
\begin{multline}\label{particle T}
    T_{\alpha\beta}^\pm = \frac{f }{\mathcal{C_\pm} E \rho^2} u_\alpha
    u_\beta \delta(\rho - \rho_p)
    \left[ \frac{1}{4\pi} + \right. \\ \left. \sum_{l=1}^\infty \sum_{m=-l}^{l} Y_{lm}(\Omega) Y^*_{lm} (\Omega_p)
    \right].
\end{multline}
Here, we have defined
\begin{equation}
    \mathcal{C}_\pm = \frac{(GM)^3}{M_p e^{ky_\pm}}, \quad E = -g_{\alpha\beta} u^\alpha
    \xi_{(t)}^\beta, \quad \xi_{(t)}^\alpha = \di_t.
\end{equation}
As usual, $E$ is the particle's energy per unit rest mass defined
with respect to the timelike Killing vector $\xi_{(t)}^\alpha$.

Comparing (\ref{Theta defn}) and (\ref{Theta decomposition}) with
(\ref{particle T}), we see that
\begin{subequations}
\begin{eqnarray}
    \dimT^{(s)}_{\alpha\beta} & = & \frac{f}{\sqrt{4\pi} E \rho^2} u_\alpha
    u_\beta \, \delta[\rho - \rho_p(\tau)], \\
    \Lambda_1 & = & \frac{f}{\sqrt{4\pi} E \rho^2} \, \delta[\rho - \rho_p(\tau)], \\
    \Lambda_2 & = & \frac{E\dot{\rho}_p}{\sqrt{4\pi} f \rho^2} \, \delta[\rho - \rho_p(\tau)], \\
    \Lambda_3 & = & \frac{f \tilde{L}^2 }{\sqrt{4\pi} E \rho^4} \delta[\rho -
    \rho_p(\tau)],
\end{eqnarray}
\end{subequations}
where $\dot{\rho}_p = d\rho_p / d\tau$. Here, we have identified $L$
as the total angular momentum of the particle (per unit rest mass),
defined by
\begin{equation}
    \frac{L^2}{r^2} = \gamma_{\alpha\beta} u^\alpha u^\beta, \quad \tilde{L} = \frac{L}{GM}.
\end{equation}
Note that for particles travelling on geodesics, $E$ and $L$ are
constants of the motion.  These are commonly re-parameterized
\cite{Cutler:1994pb} in terms of the eccentricity $e$ and the
semi-latus rectum $p$, both of which are non-negative dimensionless
numbers:
\begin{gather}
\begin{split}
    E^2 & = \frac{(p-2-2e)(p-2+2e)}{p\,(p-3-e^2)}, \\
    \tilde{L}^2 & = \frac{p^2}{p-3-e^2}.
\end{split}
\end{gather}
The orbit can then be conveniently described by the alternative
radial coordinate $\chi$, which is defined by
\begin{equation}
    \rho = \frac{p}{1+e\cos\chi}.
\end{equation}
Taking the plane of motion to be $\theta = \pi/2$, we obtain two
first order differential equations governing the trajectory
\begin{gather}
\begin{split}
    \frac{d \chi}{d \tau} &= \left[ \frac{(p-2-2e\cos\chi)^2(p-6-2e\cos\chi)}
    {\rho_p^4 (p-2-2e)(p-2+2e)}\right]^{1/2}  , \\
    \frac{d \phi}{d \tau} &= \left[ \frac{p\,
    (p-2-2e\cos\chi)^2}{\rho_p^4 (p-2-2e)(p-2+2e)} \right]^{1/2} .
\end{split}
\end{gather}
These are well-behaved thorough turning points of the trajectory
$d\rho_p/d t = 0$.  When $e < 1$ we have bound orbits such that
$p/(1+e) < \rho_p < p/(1-e)$, while for $e > 1$ we have unbound
`fly-by' orbits whose closest approach is $\rho_p = p/(1+e)$.  To
obtain orbits that cross the future event horizon of the black
string, one needs to apply a Wick rotation to the eccentricity $e
\mapsto ie$ and make the replacement $\chi \mapsto i\chi + \pi/2$.
Then a radially infalling particle corresponds to $e = \infty$.

Since this type of brane matter will be the topic of the rest of
this paper, it is worthwhile to write out the associated source
terms in the wave equation explicitly as a function of orbital
parameters
\begin{gather}
\begin{split}\nonumber
    \mathcal{S}_\psi = & \frac{2f^2 \dot{\rho}_p }{3\sqrt{4\pi}E(2+\mu^2\rho^3)}
    \Bigg[ -(2\rho^2 + 3\tilde{L}^2) \delta'[\rho - \rho_p(\tau)]
     \\ & + \frac{6\rho E^2}{f} \left( \frac{\mu^2\rho^3-4}{\mu^2\rho^3+2} \right) \delta[\rho -
     \rho_p(\tau)]
    \Bigg],
\end{split}\\
    \mathcal{S}_\varphi = -\frac{f^2 \dot{\rho}_p}{6\sqrt{4\pi}E\rho}
    \delta'[\rho - \rho_p(\tau)].\label{explicit source terms}
\end{gather}
Note that
\begin{gather}
\begin{split}
    |\dot{\rho}_p| & \, < \, f, \\
    \dot{\rho}_p = 0 \, & \Rightarrow \,
    \mathcal{S}_\psi = \mathcal{S}_\varphi = 0, \\
    E \gg 1 \, & \Rightarrow \,
    \mathcal{S}_\psi \gg \mathcal{S}_\varphi.
\end{split}
\end{gather}
That is, the particle's speed is always less than unity, the sources
wave equation vanish if the particle is stationary or in a circular
orbit, and high-energy trajectories imply that the system's dynamics
are not too sensitive to brane-bending modes $\psi \gg \varphi$.

\section{Some Typical Waveforms}\label{sec:typical}

We shall now integrate our coupled system of equations
(\ref{eq:master}) (i.e.~the master equation and the brane bending
equation) for a variety of different orbital parameters. Before we
investigate some of the typical waveforms which appear, let us
briefly digress on some issues involved in the integrations and the
resulting waveforms.

First let us consider the solution to the wave
equation,~(\ref{eq:master}), with no source, ${\cal S}=0$.  This
solution is commonly excited by taking Gaussian initial data
somewhere near the photon sphere and letting the system evolve. For
a normal black hole this results quasi-normal ringing followed by a
power law tail at late times, as seen by a distant observer. In our
case, quasi-normal ringing is subdominant and instead the signal
behaves roughly as an oscillating power-law \be \psi\sim
\tau^{-\alpha(\tau,\mu)}\sin [\omega(\tau,\mu)\tau ], \ee where
$\alpha$ and $\omega$ are slowly-varying functions of $\tau$, (as
compared to the characteristic timescale $1/\mu$). While it would
take a detailed numerical investigation to determine the precise
nature of these functions for the S-wave potential, it has been
shown~\cite{Koyama:2001ee,Burko:2004jn} that at late times
$\tau\gg1/\mu^3$ we have, independently of $\mu$, $\alpha\rightarrow
5/6$ (from above) while the frequency \emph{increases} to its
asymptotic value $\mu$ as follows: \be
\omega(\tau,\mu)\rightarrow\mu\left[1-\frac{3}{2}\left(\frac{2\pi}{\tau}\right)^{2/3}\right].
\ee

The next issue concerns our approximation for the $\delta$-functions
appearing in the source terms (\ref{explicit source terms}). These
we shall approximate by a Gaussian profile in the $x$-coordinate
\cite{LopezAleman:2003ik,Sopuerta:2005rd}:%
\begin{multline}
\delta(r-r_p)=\frac{1}{GMf}\delta(x-x_p) \\ =
\frac{1}{GMf\sqrt{\pi}\varepsilon}\exp
\left[-\frac{(x-x_p)^2}{\varepsilon^2}\right],
\end{multline}
which becomes exact in the limit $\varepsilon\rightarrow0$. Provided
that the full width at half maximum (FWHM) $2\sqrt{\ln
2}\varepsilon$ of the Gaussian is much less than the characteristic
scales we are interested in~-- namely $\mu$ and the `width' of the
potential~-- this is a good approximation (and is also why we choose
a Gaussian in $x$ and not $r$, so that the profile remains thin
inside the photon sphere).

The third involves our treatment of initial data. In normal
relativity, one switches on the interaction at some time; the
shock in the wave equations produced by this propagates way at the
speed of light. Here, however, a massive mode signal is produced
which decays very slowly. This makes it difficult to disentangle
the real signal we are interested in from this spurious signal; we
shall discuss this further as it arises.

A further issue which appears is the gravitational waves produced just by the unaccelerated motion of the particle itself. Evolving a geodesic compact source in flat space within
GR does not produce gravitational waves (at linear order). With massive modes of the graviton present, however,
this is not the case: an observer sees a wavetail after the
particle has passed with a wavelength roughly that of the massive
mode. This effect also tangles itself up in the waveforms we are actually interested in.

\begin{figure}[ht]
\includegraphics[width=\columnwidth]{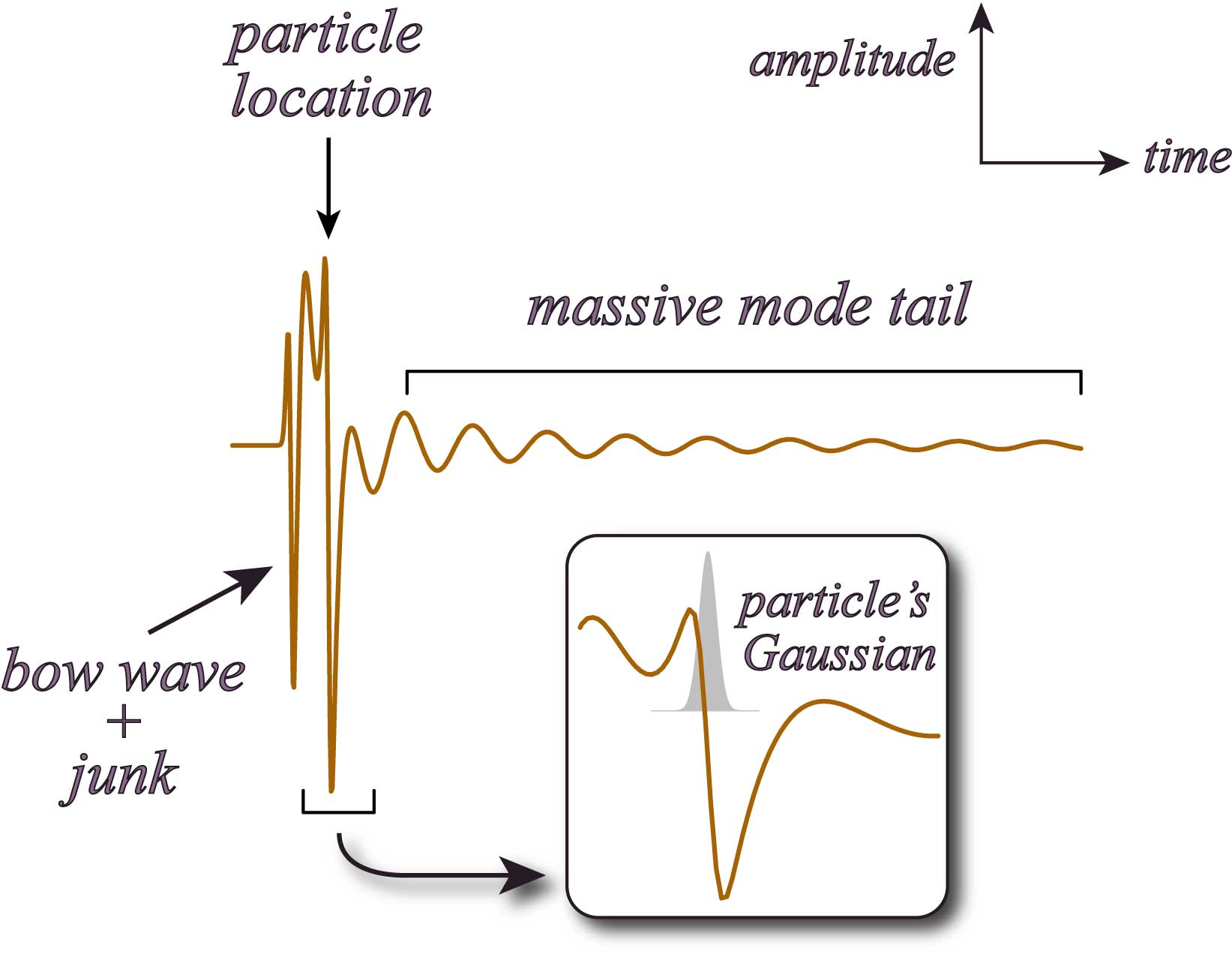}
\caption{Waveform of a particle moving past an observer in the far
field.
\label{fig:far.field}}
\end{figure}

These issues are illustrated in Fig.~\ref{fig:far.field}, where we
show $\psi$ for a particle moving in the far field as seen by an
observer located at $x=200$. The particle is on a plunge orbit with
$\epsilon=0.1$ and $p=3.09$, and we have shown the lowest mass mode,
$m_1=0.5$. The integration was started with the particle located at
$x\approx240$, and we have used $\varepsilon=1/3$.
There are three key features. The first is the bow wave which
precedes the particle: this is just junk from the initial data
which we want to minimise. This junk will not interfere with the
signal from the particle interacting with the black string,
provided we start the simulation when the particle is in the far
field: in this case, the spike from the particle increases by over
an order of magnitude by the time it gets to $x\approx30$,
dwarfing any contamination from the junk.

The second feature is the particle passing the observer: the
disturbance may be compared to the width of the Gaussian, whose FWHM
is displayed by the width of the stem of the arrow pointing to it
(and is thinner than the width of the line displaying the signal).
We can see that the disturbance length scale is much wider implying
that the Gaussian is thin enough. The main part of the signal is the
massive mode tail which exists in the particle's wake. This has a
characteristic power law decay discussed above; this part of the
signal causes problems later as typically it will not have decayed
away by the time the signal from the black string reaches the
observer, for interesting observer locations (we shall see that
interesting signals occur relatively near the black string, so this
has ramifications later).

\subsection{Plunge orbits and the hierarchy of massive modes}

We shall investigate here in some detail the situation of a plunge
orbit depicted in
Fig.~\ref{fig:S-Wave...UnBound...Orbit=[p=3.09,e=.1].....pic1.orbit.ps},
with $\epsilon=.1$ and $p=3.09$ (which corresponds to $E=2.0,
L=9.78$).
\begin{figure}[ht]
\includegraphics[width=0.8\columnwidth]{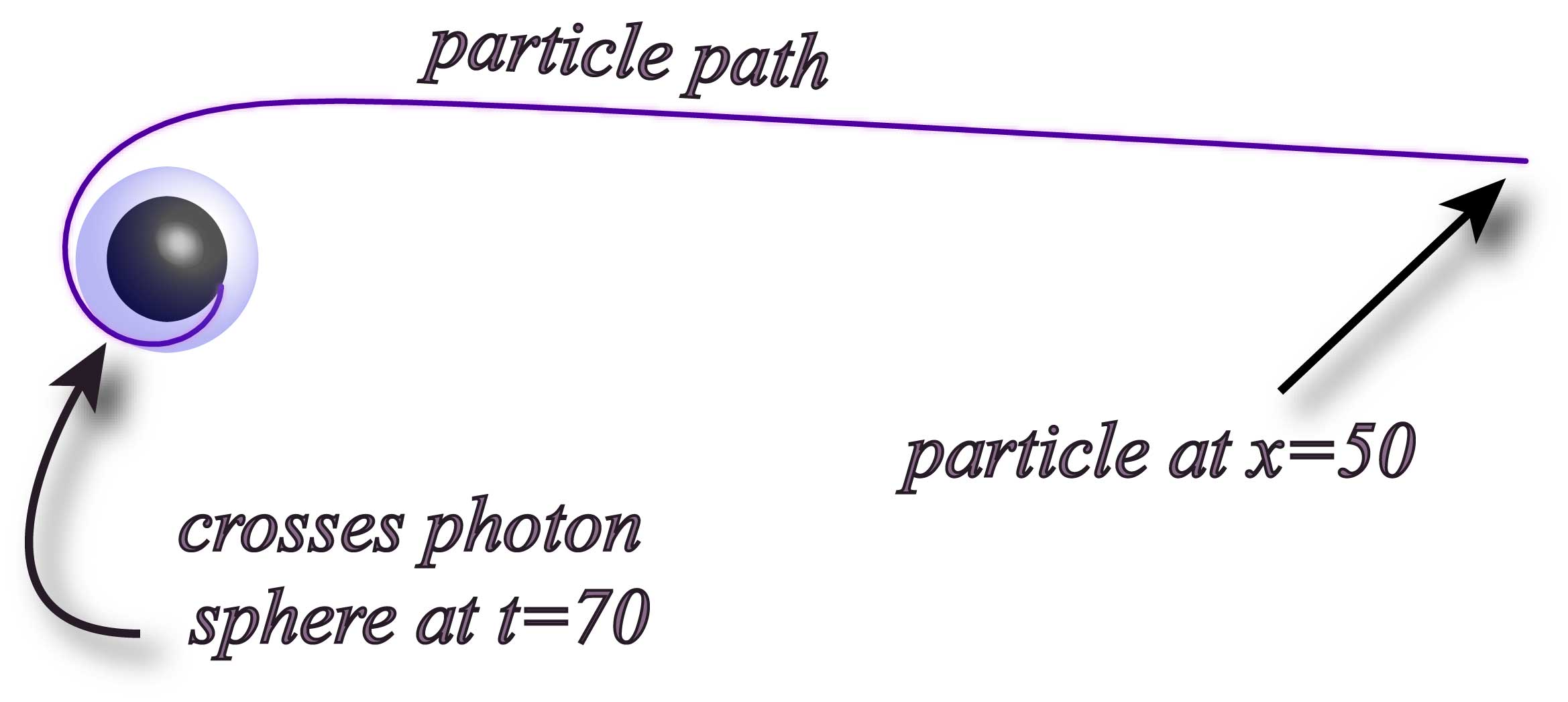}
\caption{Schematic of our plunge orbit.
\label{fig:S-Wave...UnBound...Orbit=[p=3.09,e=.1].....pic1.orbit.ps}}
\end{figure}
We shall show how the hierarchy of mass modes contribute to the
total spherical signal. Assuming we are in the ``light mode'' regime
$m_n\ell \ll 1$, the KK masses are given by \be
\mu_n=[1,1.831,2.655,3.477, \ldots]\mu_1. \ee The string of KK
masses we shall use has $\mu_1=0.5$, which starts just above the GL
instability, where $\mu_{GL}\approx0.4301$. For $d/\ell=20$ this
corresponds to a black string of mass $4.3M_\odot$, while for
$d/\ell=35$, we have a $1.4\times10^9M_\odot$ black string.  We will
present composite solutions for $\psi$; i.e.,
\begin{equation}
    \psi(t,x) = \sum_{n} (z^+_n)^2 \psi_n(t,x),
\end{equation}
where $\psi_n$ is the numeric solution for a given mass $\mu_n$ and
the $z_+$ parameters are given by (\ref{light z's}). We have assumed
that both the observer and the source are on the visible brane. Note
that if we wanted to reconstruct the full spherical GW signal, we
would first have to apply the inversion formulae (\ref{inversion
formulae}) to each of the $\psi_n$ to get $h_{\alpha\beta}^{(n,s)}$
[c.f.~(\ref{spherical h decomp})] and then sum over $n$ using
(\ref{main mode decomposition}) to obtain the spherical part of
$h_{AB}$.  However, the simplified composite signal given above will
capture most of the essential features of the complete spherical GW
signal, and will be sufficient for the qualitative discussion given
here.
%

We show, in
Fig.~\ref{fig:S-Wave...UnBound...Orbit=[p=3.09,e=.1].....pic1.signal.z=100.ps},
the composite signal $\psi$, and the brane-bending contribution
$\xi$, as seen by an observer at $x=100$ for this plunge orbit,
starting when the particle passes the observer at $\tau=0$. The
integration was started with the particle located at $x\approx240$,
so initial data problems give a very small contamination of the
signal, and we have used $\varepsilon=1/3$.
\begin{figure}[ht]
\includegraphics[width=\columnwidth]{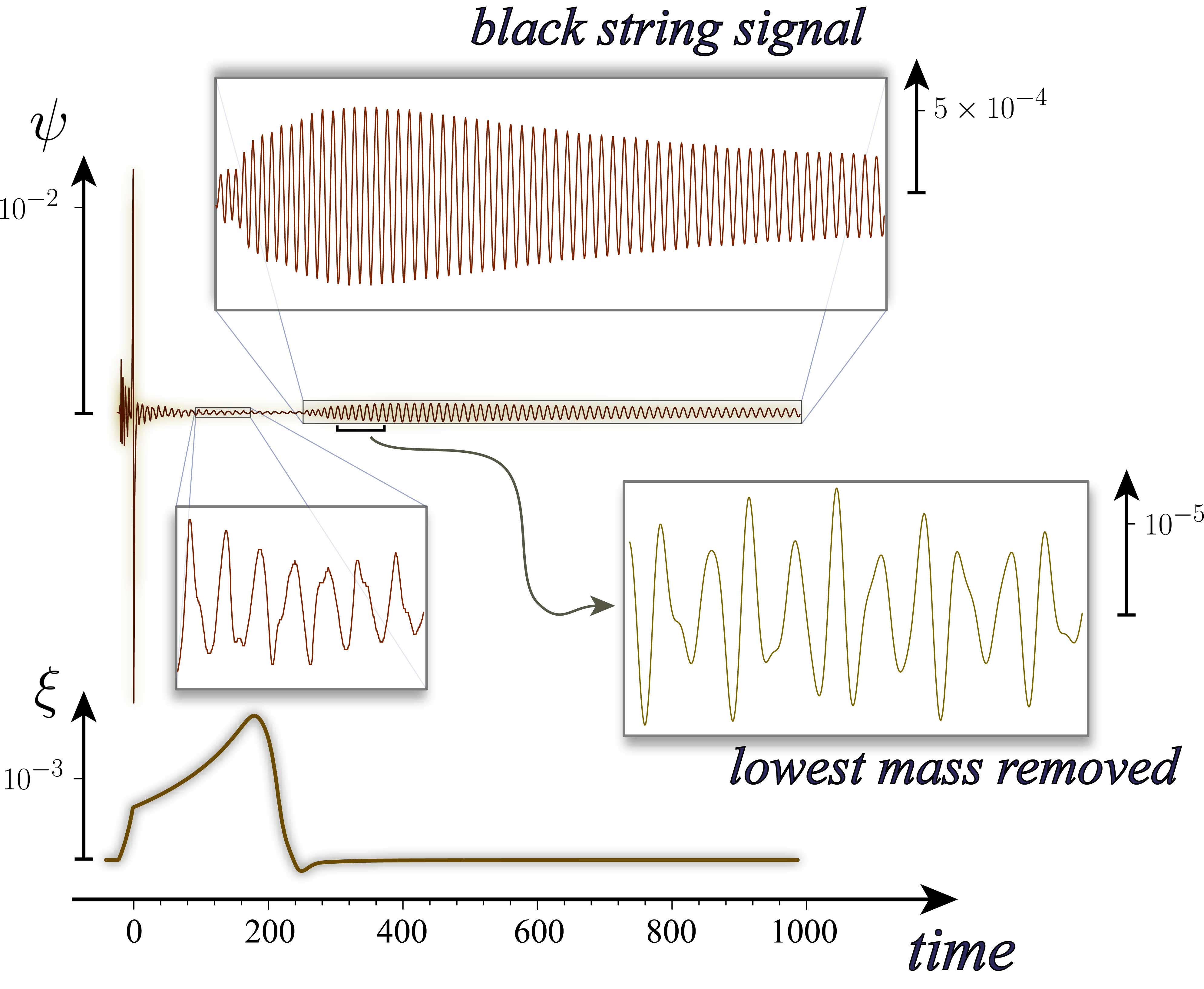}
\caption{Composite signal from a plunge orbit (top) with brane
bending contribution (bottom) for an observer at $x=100$, for the first 4 mass modes.
\label{fig:S-Wave...UnBound...Orbit=[p=3.09,e=.1].....pic1.signal.z=100.ps}}
\end{figure}

The gravity wave signal, $\psi$, has two distinct parts. The
first, as we have discussed, is from the particle itself and the
wave-tail it leaves in its wake. But now we have contributions
from higher mass modes, which give a distinct wobble to the tail,
shown in the bottom left blow-up. The second is from the black
string itself (upper blow-up). As the particle falls into the
black string it emits radiation which in the frequency domain is
sharply peaked about the frequency of the massive mode (of which
more later). Being massive, much of this radiation falls into the
string, but some of it makes it out to the observer, the first
hint of which arrives around $\tau\approx300$. This signal reaches
a peak around $\tau\approx 4-500$~-- a considerable length of time
compared to a comparable GR signal~-- and then gently turns into
the characteristic power-law tail fall-off. A key feature of this
is the lack of influence the higher mass modes have, compared to
$\mu_1$; the signal with $\mu_1$ removed is shown in the blow-up
at bottom right. We can see from the  relative scales that this is
suppressed by nearly two orders of magnitude. Compare this to the
earlier tail from when the particle passes the observer~-- massive
modes higher than $\mu_1$ are clearly visible there. The
conclusion of this is that signals arising from excitations of the
string itself are overwhelmingly dominated by the lowest mass
mode.

The overall amplitude of the excitation is worth noting:
$\psi_{\mathrm{max}}\sim 5\times 10^{-4}$. Given that the source
term from the particle is $\mathcal{O}(1)$, one might naively expect
a signal of comparable strength~-- indeed, this is roughly what
happens in GR. Such a weak excitation clearly indicates that
spherical massive modes are only weakly stimulated by the particle
in-fall. In part this is due to the fact that some of the signal
falls into the string; more on this later.

Finally, we come to the composite brane-bending contribution to
the signal. The signal, which is independent of $\mu$, is pretty
featureless.  As the particle passes, a dent in the brane
accompanies it; this reaches a peak after the particle has passed,
and slowly relaxes back to zero without oscillating. As the brane
remains significantly bent long after the particle passes, this
extends the total source feeding the gravity wave signal beyond
the particle's Gaussian. Thus, the black string gets a far longer
stimulation than it would otherwise get from a point source: the
brane bending signal is partly responsible for the length of time
$\psi$ remains peaked in the latter part of the signal. This may
be seen by the fact that the tail part of the signal has a
power-law fall off of $\alpha \approx 1.1$ at $\tau\approx1000$,
so hasn't yet reached the asymptotic late time value of $5/6$.
However, comparison of the signal with the brane bending switched
off reveals that the contribution to the amplitude is only of the
order of a few percent.

\subsection{Bound flower-shaped orbits: steady state waveforms}

\begin{figure}[ht]
\includegraphics[width=0.8\columnwidth]{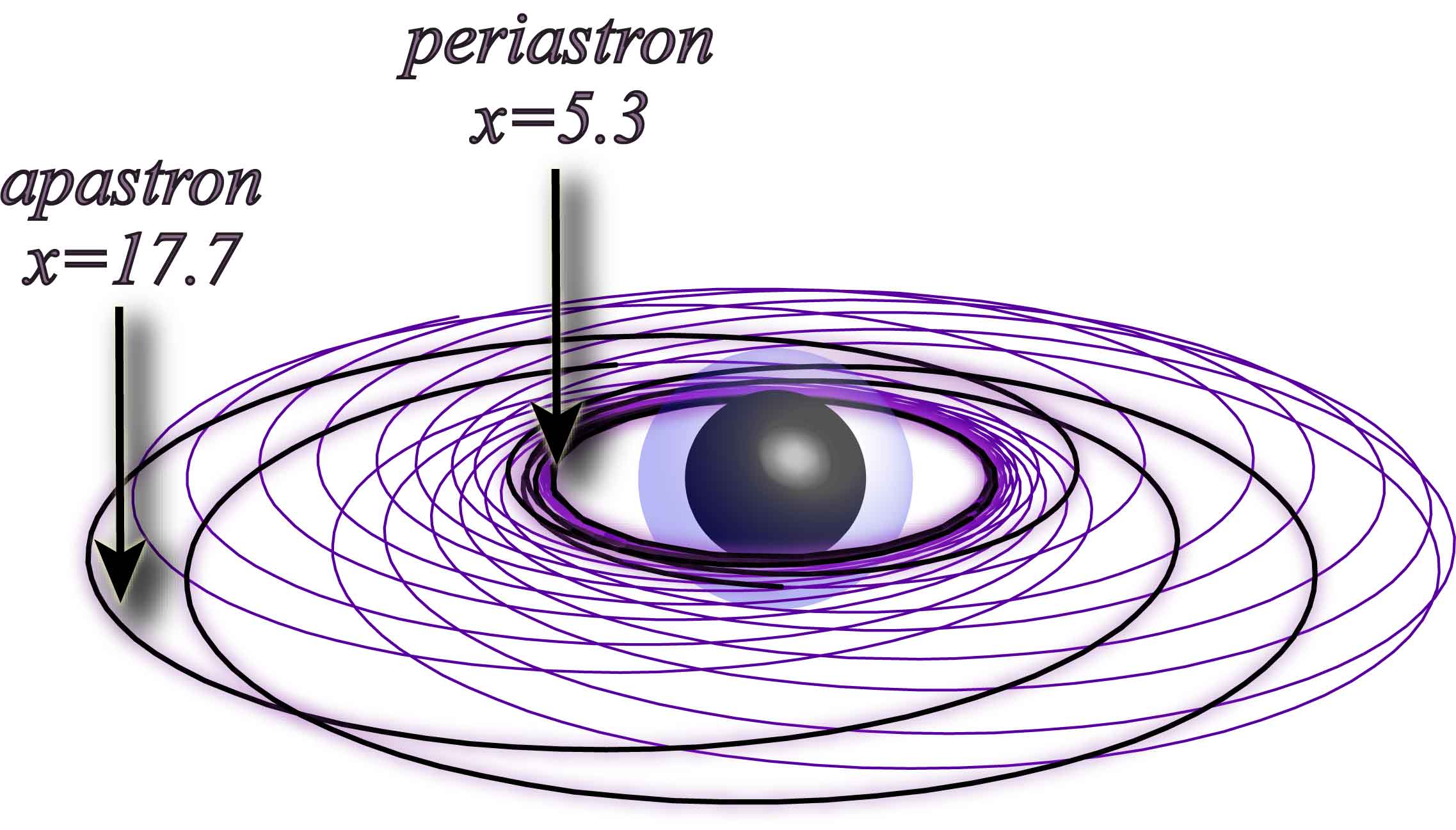}
\caption{Illustration of the  flower-shaped bound orbit used for
investigating the steady state waveforms
\label{fig:flower.orbit.ps}}
\end{figure}

Let us now investigate the signal which comes from a bound orbit,
illustrated in Fig.~\ref{fig:flower.orbit.ps}, and explore how the
signal changes with distance from the black string. We concluded
from the previous section that the higher mass modes add only a
small correction to the full signal, so here we shall only
investigate the signal arising from the lowest mass, $\mu_1=0.5$.
We choose an orbit with $e=0.5, p=7.05$, and we take
$\varepsilon=1$.

Integration of the equations is complicated by initial data, once
again, but more-so than in the plunge case. This is because the
amplitude of the source in the wave equations increases with
decreasing $x$, so the best we can do is start the integration at
the apastron where it's smallest, and wait for the contamination
to pass the observer. Unfortunately the wave tail makes this quite
a long time~-- roughly $\tau \sim 4-5000$ for an observer around
$x\lesssim 100$, compared to $\tau\sim100$ in GR. After this time
the desired steady-state waveform is reached, which we set to
$\tau=0$.

\begin{figure*}[ht]
\includegraphics[width=\textwidth]{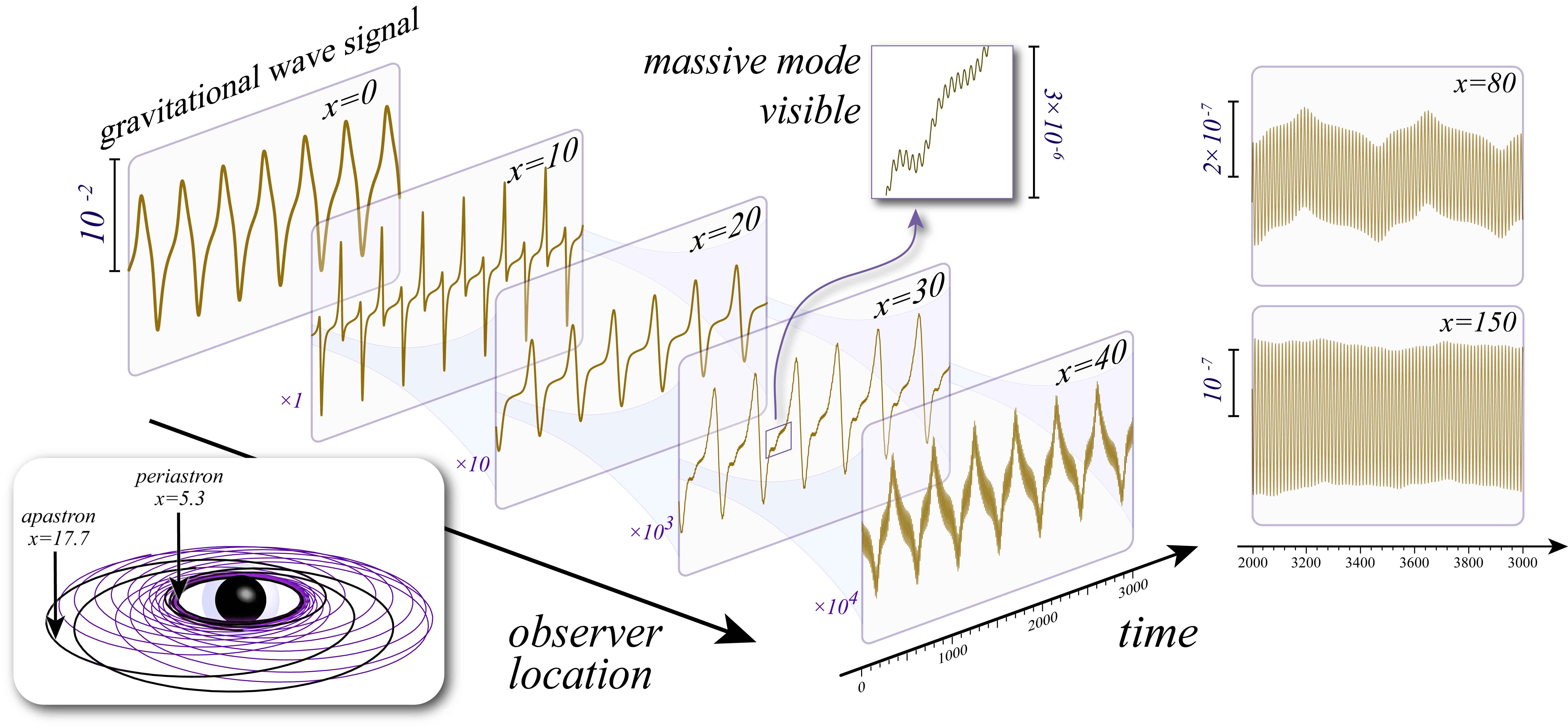}
\caption{The steady-state gravity wave signal $\psi$ as seen by
successively distant observers from the black string, for the lowest mass mode.
\label{fig:flower.orbit.figuresHR.ps}}
\end{figure*}

In Fig.~\ref{fig:flower.orbit.figuresHR.ps} we show the results of
this integration for several observer locations.
As the source evolves it oscillates along the $x$ direction,
between about $5.3$ and $17.7$.  It can be immediately seen that
observers see radically differing signals depending on whether
they are in the near, intermediate, or far zones.

\begin{description}

\item[Near Zone:] An observer sitting near the photon sphere around $x=0$ will
see a relatively normal signal: a gravity wave propagating at light
speed (since $V\approx0$ there) which precisely mimics the behaviour
of the source. Around $x=10$~-- located in the `middle' of the
orbit~-- the amplitude of $\psi$ reflects the source passing back
and forth.

\item[Intermediate Zone:] Here things get a bit more interesting. As we move further out
to $x\sim40$ the direct orbital signal gets damped dramatically~--
by 4 orders of magnitude. This is due to low frequency parts of the
signal being exponentially damped as they travel through the
potential (all frequencies roughly less than $\mu$ get damped). By
$x\sim30$ the massive modes become visible as a high frequency
wobble riding on the orbital part. And around $x\sim40$ the two
components become equally dominant.

\item[Far Zone:] As we observe from more distant locations where the potential
is almost flat virtually all the low frequency components have been
suppressed, and we are left with a low amplitude massive signal. A
gentle oscillation to the envelope is all that remans of the orbital
signal.

\end{description}

One of the interesting and unexpected things about this is that
the massive modes are excited at all given that the frequency of
the source is orders of magnitudes smaller than the mass. Also
note that the amplitude of the signal falling into the black
string is about five orders of magnitude larger than the signal
which makes it out.

\subsection{Flyby orbits}

\begin{figure}[ht]
\includegraphics[width=\columnwidth]{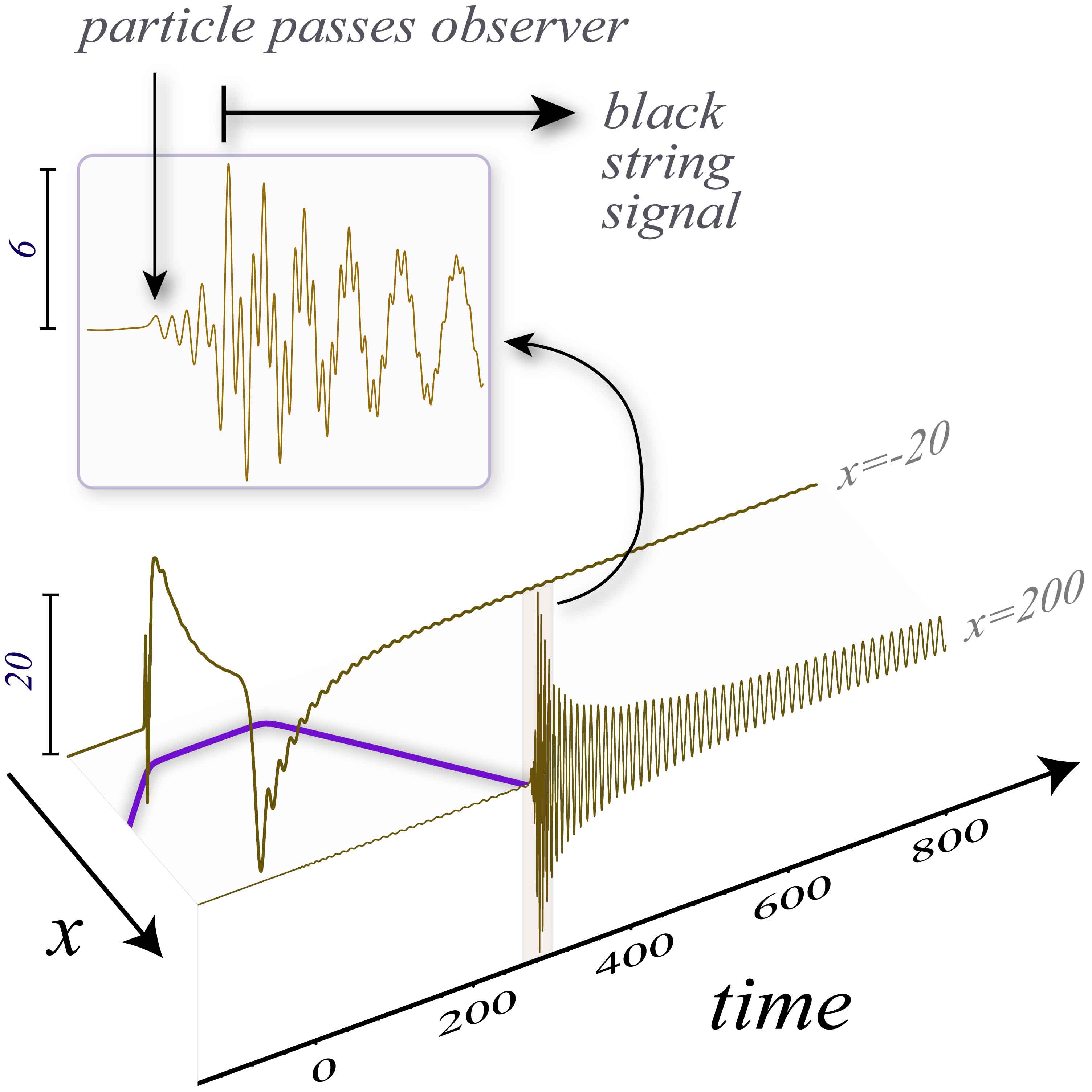}
\caption{The GW signal from a fly-by orbit, for the lowest mass mode.
\label{fig:fly-by.orbit...figures.2.ps}}
\end{figure}

Finally we shall consider the case of unbound orbits with high
angular momentum. We choose an orbit with $e=3.0$ and
$p=12.00001$ (see Fig.~{fig:fly}). This completes about four orbits of the black
string, very nearly touching the photon sphere at is periastron
($\tau=0$ in this simulation).
\begin{figure}[ht]
\includegraphics[width=0.5\columnwidth]{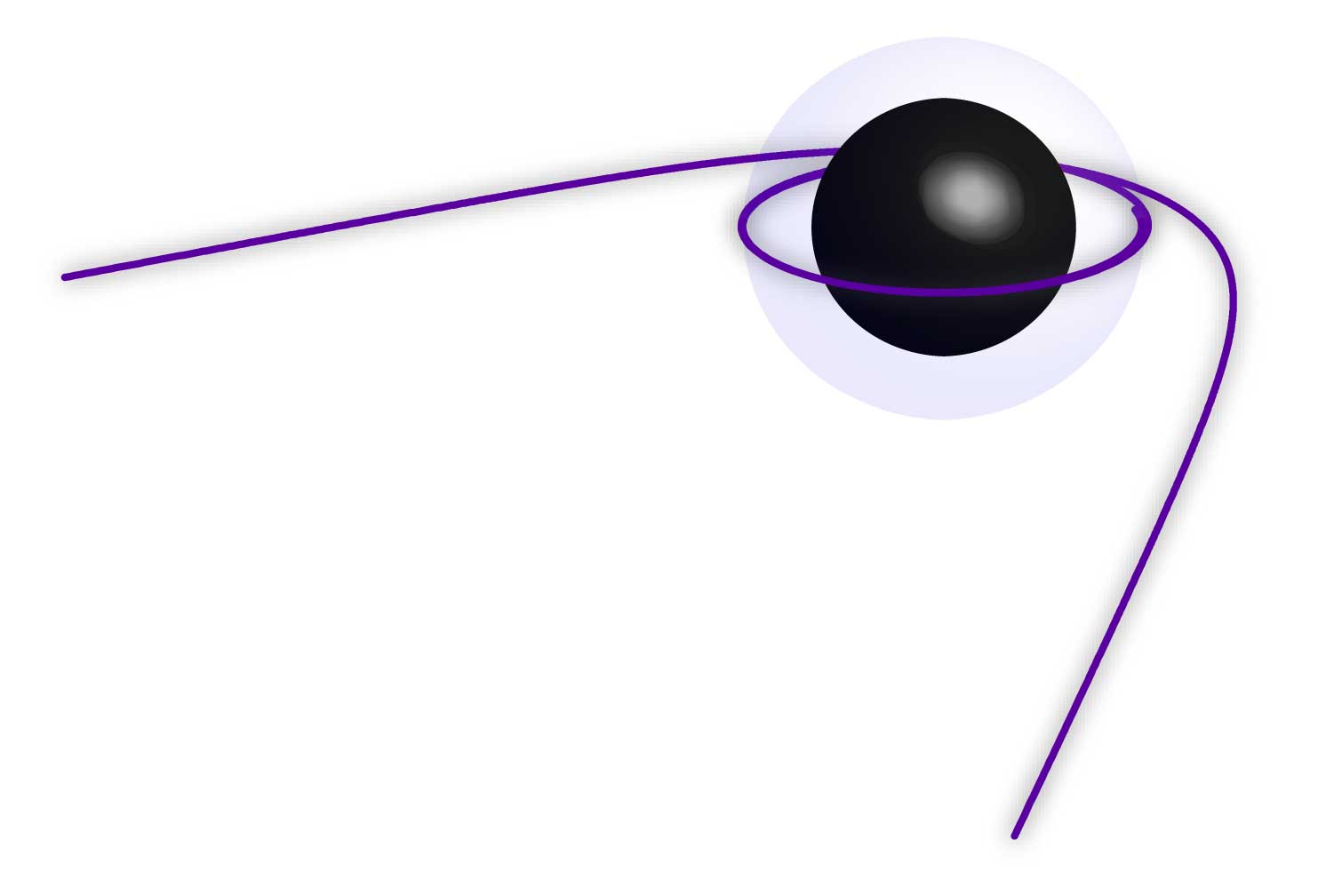}
\caption{Schematic of the fly-by orbit.
\label{fig:fly}}
\end{figure}
 If we think of this orbit in terms
of the $x$ coordinate, the particle comes in on a nearly null ray,
decelerates very rapidly to zero at the photon sphere where it
lingers for $\tau\sim 150$. It then rapidly accelerates off to
infinity (of course the particle is geodesic so experiences no
acceleration). The source in the wave equation becomes very large
at these two events. We see in
Fig.~\ref{fig:fly-by.orbit...figures.2.ps} how this acceleration
can induce a strong GW signal.
First consider the GW which falls into the string. This consists
of two pulses corresponding to these two accelerations. As the
potential is basically flat in this zone the pulses reflect
closely the time evolution of the source. We can see some massive
modes on top of this caused by a reflection off the potential.

In the far zone, on the other hand, only the second acceleration
produces a GW signal~-- and this is of nearly the same amplitude
as the signal which falls into the string. In the previous two
examples the far zone signal was orders of magnitude smaller than
that which passes the horizon. The burst of GW seen at $x=200$
nearly coincides with the passing of the particle, although the
blow-up shows the the particle precedes the signal somewhat. The
peak part of the signal has some interesting wobbles; after this
the waveform takes its familiar shape of a decaying massive mode.

The interesting aspect of this simulation is that the amplitude is
orders of magnitude larger than the previous two cases. We can
understand why this happens by examining the source term, given by
Eqs.~(\ref{explicit source terms}). The two terms which are most
important are the coefficients of $\delta'[\rho-\rho_p(\tau)]$. This
is proportional to $u^r$, which when $E$ and $L$ are large scale
like $(p-3-e^2)^{-1/2}$ ($=10^{2}$ for the simulation above); thus
when $p\rightarrow3+e^2$ we have an infinite source and signal. Note
that $u^r=0$ at the periastron, so the source peaks near the
periastron, when $r$ is small, but not far away when $1/r$ terms
kick in. We have numerically confirmed that the signal which makes
it to infinity does in fact scale in this way.

\section{Discussion}

In this paper, we have presented the derivation and numeric solution
of the equations of motion for gravitational waves in the black
string braneworld sourced by brane localized matter.  In
\S\ref{sec:RS}, we presented a generalized two brane Randall-Sundrum
model and then specialized to the black string background.  In
\S\ref{sec:linear}, we considered the linear perturbations of the
model and the introduced the Kaluza-Klein massive mode decomposition
\S\ref{sec:KK}.  The limit under which the model reduces to
Brans-Dicke theory was discussed in \S\ref{sec:brans-dicke}, which
led to a constraint on the brane separation.  We discussed the
specialization of the formalism to spherical radiation
(\S\ref{sec:spherical}) and pointlike sources (\S\ref{sec:point}).
Finally, in \S\ref{sec:typical} we presented the results of numeric
simulations of the spherical GWs produced by perturbing bodies
undergoing plunge, bound, and fly-by orbits.

Future work on this model involves improving our simulation
techniques by incorporating characteristic integration techniques
\cite{Seahra:2006tm} that more naturally deal with the
delta-functions in the GW source.  It would also be interesting to
consider more realistic modeling of sources of finite size.  Once
this is accomplished we can build up a bank of simulations for a
variety of orbital parameters, choices of $\mu$, and other
multipoles. One can then systematically begin looking for these
waveform in the data obtained from gravitational wave detectors, and
thereby provide a means of further constraining the Randall-Sundrum
braneworld model.

\begin{acknowledgments}

We thank Roy Maartens for discussions. SSS is supported by NSERC
(Canada).  CC is supported by the NRF (South Africa).

\end{acknowledgments}

\bibliography{s-wave}

\end{document}